\documentclass[useAMS,usegraphicx]{mn2e}

\def\iron{\hbox{[Fe\,{\sc ii}]}}
\def\hii{\hbox{H\,{\sc ii}}}

\def\he{\hbox{He\,{\sc i}}}
\def\sivi{\hbox{[Si\,{\sc vi}]}}
\def\sivii{\hbox{[Si\,{\sc vii}]}}
\def\caviii{\hbox{[Ca\,{\sc viii}]}}
\def\oiii{\hbox{[O\,{\sc iii}]}}

\def\oiii{\hbox{[O\,{\sc iii}]}}

\def\alix{\hbox{[Al\,{\sc ix}]}}

\def\PM{$\pm$}
\def\kms{km s$^{-1}$}

\def\brg{Br$\gamma$}

\def\ergs{ergs cm$^{-2}$ s$^{-1}$}
\def\h2{H$_2$}

\newif\ifAMStwofonts
\AMStwofontstrue

\title[NIR spectroscopy of nearby Seyferts. II.]{Near-infrared spectroscopy 
  of nearby Seyfert galaxies. II. Molecular content and coronal 
  emission}
\author[J. Reunanen et al.]{J. Reunanen$^1$, J.K. Kotilainen$^1$, M.A. 
  Prieto$^{2,3}$\\
  $^1$ Tuorla Observatory, University of Turku, V\"ais\"al\"antie 20, 
  FIN--21500 Piikki\"o, Finland; jupere@utu.fi, jarkot@utu.fi\\
  $^2$ European Southern Observatory, Karl-Schwarzschild-Str. 2, D--85748 
  Garching bei M\"unchen, Germany\\
  $^3$ Max-Planck-Institut f\"ur Astronomie, K\"onigstuhl 17, D-69117 Heidelberg, Germany; prieto@mpia-hd.mpg.de}
\date{Accepted date +  Received date }
\pagerange{\pageref{firstpage}--\pageref{lastpage}}
\pubyear{2002}

\begin{document}
\label{firstpage}

\maketitle

\begin{abstract}
  
  We present sub-arcsec near-infrared 1.5 -- 2.5 $\mu$m moderate
  resolution long-slit spectra of eight nearby Seyfert galaxies
  ($z<$0.01), both parallel to the ionization cone and perpendicular
  to it. These spectra complement similar data on six Seyferts,
  presented in Reunanen, Kotilainen \& Prieto (2002), and are used to
  study the spatial extent of the line emission, integrated masses of
  excited {\h2} and the excitation mechanisms of interstellar gas.
  
  Large concentrations of molecular gas ({\h2}) are present in the
  nucleus regardless of the Seyfert type. The spatial extent of the
  \h2 emission is larger perpendicular to the cone than parallel to it
  in 6/8 (75 \%) galaxies, in agreement with the unified models of
  Active Galactic Nuclei. The FWHM sizes of the nuclear {\h2} emission
  range from $<$20 pc to $\sim$300 pc, and are larger than the
  predicted sizes for molecular torus (1-100 pc). Thus the emission
  likely arises from the material surrounding the torus rather than directly
  from the torus.
  
  Broad Br$\gamma$ was detected in nearly half of the optically
  classified Seyfert 2 galaxies, including two objects with no evidence
  for hidden polarized Broad Line Region. This high
  detection rate stresses the importance of extinction effects as the
  main cause for the Seyfert dichotomy.
  
  Br$\gamma$ and {\iron} correlate both spatially and kinematically.
  Nuclear {\iron} emission is generally blueshifted which together
  with high {\brg}/{\iron} ratios suggests shocks as the dominant
  excitation mechanism in Seyfert galaxies.
  
  Bright coronal emission lines {\sivi} and {\sivii} are common in
  Seyferts, as they are detected in $\sim$60 \% of the galaxies. In
  three galaxies the coronal lines are extended only in the direction
  parallel to the cone. This could be explained by a strongly
  collimated radiation field or, most plausibly, by shock excitation
  due to the jet or superwind interacting with the interstellar
  medium.

\end{abstract}


\section{Introduction}

In the currently favoured unified models of Active Galactic Nuclei
(AGN; e.g. Antonucci 1993), a thick molecular torus surrounds the
nucleus.  In Seyfert 1 (Sy1) galaxies the nucleus and the Broad Line
Region (BLR) are visible, while in Seyfert 2 (Sy2) galaxies the torus
obscures the nucleus and the BLR and only the Narrow Line Region (NLR)
is visible. Evidence for the unified models includes the detection of
polarized broad lines in many Sy2s (e.g. Moran et al. 2000), believed
to be light from the obscured BLR reflected into our line-of-sight,
and cone-like {\oiii} structures (e.g. Mulchaey, Wilson \& Tsvetanov
1996), almost invariably aligned with the radio jets.

The predicted size of the molecular torus in AGN ranges from 1 to 100
pc (e.g. Pier \& Krolik 1993).  Adequate spatial resolution to search
for material associated with the torus, with a minimized contamination
from circumnuclear star formation, can be achieved in the
near-infrared (NIR; e.g. Veilleux, Goodrich \& Hill 1997), where a
wealth of emission and absorption lines are available. However, until
recently such NIR spectroscopic studies have been made at only
moderate spatial resolution (Veilleux et al. 1997: 3 arcsec aperture,
$R=1400$; Winge et al 2000: 7--8 arcsec, $R=250$; Sosa-Brito et
al. 2001: 2 arcsec, $R=1100$), and/or with only one slit position
angle (PA) along the radio axis or the major axis of the {\oiii}
emission (e.g. Storchi-Bergmann et al.  1999), and thus have not been
able to derive the geometry of the molecular emission.

In Reunanen, Kotilainen \& Prieto (2002; hereafter Paper I), we presented 
long-slit 1.5--2.5 $\mu$m moderate resolution spectroscopy 
of the first six galaxies in our sample of nearby (0.0018 $<$ $z$ $<$ 0.010) 
Seyfert galaxies of both types with an ionization cone. 
In this paper, we complement these results with similar data on eight 
additional Seyferts, thus the size of the final sample is 14 Seyferts. 
These data are used to trace the spatial distribution, dynamics and 
excitation of molecular and nebular gas, and to determine the mass of hot 
($T > 1000$ K) molecular gas in the central regions. 
In forthcoming publications, we shall study the stellar populations and 
star forming histories of the sample, in comparison with non-Seyfert spirals.

This paper is organized as follows: In Section 2 the observations, 
data reduction and methods used in the analysis are briefly summarized.  
In Section 3 the results for individual galaxies are discussed, 
and in Section 4 we present the conclusions based on the full sample of 
14 galaxies.  Throughout this paper, 
$H_0$ = 75 {\kms} Mpc$^{-1}$ and $q_0$ = 0.5 are assumed.

\section{Observations, data reduction and methods of analysis}

The observations and data reduction are described in detail in Paper I. 
Here only a brief summary is given. Eight nearby Seyferts were 
observed in December 2000 with the 3.5 m ESO New Technology Telescope (NTT), 
using the 1024$\times$1024 px SOFI camera (Lidman, Cuby \& Vanzi 2000) 
with pixel scale 0.288 arcsec px$^{-1}$. The average seeing during the 
observations was $\sim$0.9 arcsec FWHM. The spectra were taken with the 
red grism with resolution $R = 980$ and slit width 1.0 arcsec. 
The wavelength range covered is 1.5--2.5 $\mu$m and the useful slit length 
$\sim$2 arcmin.  Short broad-band $JHKs$ images were also obtained, 
except for NGC 1068 and NGC 3783, for which we used the images from 
Peletier et al. (1999) and Mulchaey, Regan \& Kundu (1997), respectively. 
The properties of the galaxies (redshift, scale, inclination, PA of the 
major axis, morphology, AGN type), the PAs of the slit, 
total integration time and the spatial 
resolution of the $K$-band spectra are given in Table \ref{obsprop}.

\begin{table*}
  \centering
  \begin{minipage}{180mm}
    \caption{Observational properties of the galaxies. 
      \label{obsprop}}
    \begin{tabular}{llllllllllllll}
      \hline
      Galaxy     & $z$     &Scale&$i$  &PA  &Morphology&Type &PA$_\parallel$ &PA$_\perp$ &$t_{\rmn{int}}$&Resolution\\
      &            &pc/\arcsec&\degr &\degr &    &         &\degr   &\degr&min     & \arcsec\\\\
this work: & & & & & & & & & & \\
      NGC 1068   &0.00379& 75 & 32& 70 &(R)SA(rs)b       &Sy2  &12   &-78    &32,32 &0\farcs9\\
      NGC 1365   &0.00546&108 & 57& 32 &(R')SB(s)b       &Sy1.8&-55  &35     &48,48 &0\farcs7\\
      NGC 2110   &0.00779&154 & 42&157 &SAB0$^-$         &Sy2  &-3   &87     &48,48 &0\farcs9\\
      ESO 428-G14&0.00544&108 & 53&135 &SA:(l)0$^+$      &Sy2  &-50  &40     &32,32 &0\farcs8\\
      NGC 2992   &0.00771&152 & 72& 22 &Sa pec           &Sy2  &-60  &30     &64,56 &0\farcs8\\
      NGC 3081   &0.00795&157 & 39&158 &(R$_1$)SAB(r)0/a &Sy2  &-20  &70     &32,32 &1\farcs0\\
      NGC 3783   &0.00973&192 & 27&160 &(R')SB(r)a       &Sy1  &75   &-15    &48,40 &0\farcs9\\
      NGC 7582   &0.00540&107 & 65&157 &(R'$_1$)SB(s)ab  &Sy2  &64   &-26    &56,40 &0\farcs8\\\\ 
Paper I: & & & & & & & & & & \\
      NGC 1097   &0.00425&82  &46 &141 &(R'$_1$)SB(r'l)b &Sy1  &15,54&-36,-75&2$\times$32,2$\times$48&1\farcs0\\
      NGC 1386   &0.00289&56  &74 &25  &SB(s)0$^+$       &Sy2  &1    &-89    &48,64 &1\farcs1\\
      NGC 1566   &0.00483&94  &28 &30  &(R'$_1$)SAB(rs)bc&Sy1  &-42  & 48    &48,64 &1\farcs0\\
      NGC 3227   &0.00382&74  &56 &158 &SAB(s) pec       &Sy1.5&15   &-75    &48,80 &1\farcs0\\
      NGC 4945   &0.00194&19  &78 &43  &SB(s)cd: sp      &Sy2  &-64  & 26    &32,64 &0\farcs9\\
      NGC 5128   &0.00179&19  &65 &30  &S0 pec           &Sy2  &50   &-40    &16,48 &0\farcs9\\
      \hline 
    \end{tabular}
  \end{minipage}
\end{table*}

The long-slit spectra were taken both parallel and perpendicular to
the cone, in two positions along the slit separated by $\sim$2 arcmin.
Data reduction using {\sc IRAF}\footnote{ {\sc IRAF} is distributed by
the National Optical Astronomy Observatories, which are operated by
the Association of Universities for Research in Astronomy, Inc., under
cooperative agreement with the National Science Foundation} included
sky subtraction, flatfielding using dome flats, bad pixel and cosmic
ray removal, and wavelength calibration from OH sky lines and Xe arc
lamp. The spectra were averaged to remove residual continuum and OH
sky lines, divided by an atmospheric standard star, and flux
calibrated.  All the spectra of a given galaxy were combined and the
central 1.5 arcsec (5 px) spectrum was extracted, within an effective
aperture of 1.4 arcsec diameter. The emission line fluxes and
equivalent widths in this aperture are given in Table \ref{lines}. The
quoted errors are generally dominated by the continuum fitting, and
are 1$\sigma$ except 3$\sigma$ for upper limits. Because of telluric
residuals, the uncertainties of 1--0 S(3) and {\sivi} are larger,
20--40 \% of the flux.  In order to trace the extended emission, 3
pixels were usually binned together. For the faintest emission, even
larger binsizes were used.

\begin{table*}
  \begin{minipage}{170mm}
    \caption{Observed emission line fluxes within the nuclear 1.4 arcsec 
      diameter aperture, after adding all the data from different slit
      PAs. The fluxes ({\em first row}) are in units of 10$^{-15}$ erg
      cm$^{-2}$ and the Gaussian FWHMs ({\em second row}) in \AA. The
      width of the lines has not been corrected for instrumental
      profile (30--32 {\AA} depending on wavelength). Decomposition
      into two components: ({\em a}) 28\PM4 (5\% of the total flux)
      and 163\PM6 {\AA}, ({\em b}) 33\PM3 (10\%) and 178\PM20 {\AA},
      ({\em c}) 42.6\PM4 (6\%) and 185\PM6 {\AA}, ({\em d}) 47\PM10
      (7\%) and 222\PM7 {\AA}, ({\em e}) 33.9\PM1.1 (9\%) and 205\PM5
      {\AA}
      \label{lines}}
    \begin{tabular}{llllllllll}
      \hline
      Line     &$\lambda$& NGC1068  &NGC1365     &NGC2110    &ESO428-G14  &NGC2992     & NGC3081      & NGC3783    & NGC7582   \\
      Nucleus   & $\mu$m & Sy2      & Sy1.8      &Sy2        &Sy2         & Sy2        & Sy2          & Sy1        & Sy2      \\\\
      A$_K$     &        & 2.49     &1.14        &0.85       &0.29        & 0.95       & 0.28         & 1.25       & 1.51       \\\\
      $[$FeII$]$& 1.644  &29.1\PM3.2&4.1\PM0.4   &23.4\PM0.2 &7.9\PM0.1   &11.8\PM0.3  & 0.5\PM0.1    &2.6\PM0.3   & 7.1\PM0.4  \\  
                &        &49.8\PM4.5&46.5\PM8.2  &42.1\PM0.3 &37.1\PM0.4  &36.8\PM0.5  & 25.5\PM2.7   &33.5\PM3.4  & 30.2\PM2.0 \\
      1--0 S(3) & 1.958  &...       &...         &6.2        &3.1         &6.5         & 3.2          &2.2         & ...        \\
                &        &          &            &36.0       &33.7        &30.0        & 43.1         &37.4        &            \\
      $[$SiVI$]$& 1.964  &299       &...         &...        & 9.4        &7.8         & 8.1          &7.7         & 9.8        \\
                &        &84.3      &            &           &49.3        &41.0        & 36.1         &42.0        & 58         \\
      1--0 S(2) & 2.034  &$<$7      &$<$0.8      &2.4 \PM0.3 &1.1\PM0.1   &1.9\PM0.1   & 0.9\PM0.1    &$<$0.4      & 2.0\PM0.4  \\
                &        &          &            &44.1\PM4.8 &38.6\PM1.7  &32.2\PM1.8  & 37.4\PM2.2   &            & 35.6\PM8.0 \\
      $[$AlIX$]$& 2.043  &$<$7      &$<$1.0      &$<$0.3     &$<$0.2      &1.2\PM0.2   & 0.5\PM0.1    & $<$1.6     & $<$0.8     \\
                &        &          &            &           &            &33.8\PM4.6  & 33.1\PM4.8   &            &            \\
      HeI       & 2.058  &$<$7      &6.0\PM0.9   &$<$0.3     &1.0\PM0.1   &8.4\PM0.7   & 0.4\PM0.1    &7.9\PM1.5   & $<$0.7     \\
                &        &          &68\PM13     &           &35.2\PM2.6  &{\em a}     & 29.7\PM5.1   &255\PM33    &            \\
      1--0 S(1) & 2.122  &13.2\PM4.0 &$<$0.8     &4.4\PM0.3  &3.1\PM0.2   &6.5\PM0.1   & 1.6\PM0.1    &1.1\PM0.3   & 3.8\PM0.5  \\
                &        &36.9\PM8.5&            &40.4\PM1.3 &40.3\PM1.1  &33.6\PM0.4  & 33.8\PM1.2   &37.8\PM9.6  & 45.4\PM3.9 \\
      Br$\gamma$& 2.166  &34.6\PM4.1 &23.2\PM0.8 &4.8\PM0.7  &3.7\PM0.1   &23.7\PM0.5  & 1.0\PM0.1    &25.5\PM1.2  & 34.9\PM1.2 \\
                &        &74.3\PM8.9&90.3\PM3.5  &{\em b}    &44.6\PM0.7  &{\em c}     & 29.3\PM1.8   &{\em d}     & {\em e}   \\
      1--0 S(0) & 2.224  &$<$9      &$<$0.8      &1.5\PM0.3  &0.7\PM0.1   &1.5\PM0.2   & $<$0.3       &$<$0.4      & $<$0.7    \\
                &        &          &            &42.1\PM9.7 &26.1\PM2.9  &33.0\PM2.6  &              &            &           \\
      2--1 S(1) & 2.248  &$<$8      &$<$0.9      &1.2\PM0.2  &0.5\PM0.1   &1.2\PM0.2   & $<$0.2       &$<$0.4      & $<$0.8    \\
                &        &          &            &53.7\PM6.1 &39.6\PM3.3  &31.6\PM3.4  &              &            &           \\
      $[$CaVIII$]$&2.321 &67\PM9    &$<$1.3      &$<$0.5     &$<0.45$     &$<$1.5      & 1.2\PM0.2    &23.9\PM0.6  & $<$1.1    \\ 
                &        &87\PM12   &            &           &            &            & 57.3\PM8.4   &38.3\PM7.7  &           \\
      1--0 Q(1) & 2.407  &$<$22     &$<$1.5      &3.3\PM0.4  &$<$0.8      &3.6\PM0.2   & 0.7\PM0.2    &$<$0.8      & $<$1.5    \\
                &        &          &            &39.6\PM4.4 &            &28.2\PM1.4  & 19.9\PM4.7   &            &           \\
      1--0 Q(3) & 2.424  &$<$37     &$<$1.8      &$<$1.4     &1.4\PM0.4   &$<$0.3      & 1.4\PM0.1    &$<$1.1      & $<$1.9    \\
                &        &          &            &           &26.4\PM7.2  &            & 35.1\PM2.1   &            &           \\
      $[$SiVII$]$&2.483  &467\PM75  &$<$3.3      &$<$4.2     &6.6\PM1     &16\PM5      & 7.6\PM0.8    &15.3\PM3.1  &21.6\PM5.4 \\
                &        &112\PM18  &            &           &36.3\PM2.9  &110\PM30    & 48.2\PM3.9   &44\PM10     &47.8\PM9.8 \\
      \hline
    \end{tabular}
  \end{minipage}
\end{table*}

Extinction was estimated by comparing the integrated colours from the
spectra with the colours of normal, unobscured spiral galaxies
($H$-$K$ = 0.22; Hunt et al. 1997). Although this continuum extinction
may not be correct for the line emission, the difference is unlikely
to be large, as discussed in Paper I. The effect of including the
nuclear emission lines in the continuum colors is negligible, up to
$A_{\rmn{K}} = 0.02$ in NGC 2992. We have assumed the extinction law
${A_\lambda} \propto \lambda^{-1.85}$ from Landini et al. (1984) and a
foreground dust screen. The extinction for the central region is given
in Table \ref{lines} and in the spatial profile figures.

Possible mechanisms for the {\h2} emission in galaxies are thermal
(collisional) excitation by shocks (e.g. Hollenbach \& McKee 1989), UV
radiation in dense clouds (e.g. Sternberg \& Dalgarno 1989) or X-rays
(e.g. Gredel \& Dalgarno 1995), and UV pumping (fluorescence; e.g.
Black \& van Dishoeck 1987). The emission line ratios can be used to
distinguish between these excitation mechanisms. For example, the
H$_2$ line ratio 2--1 S(1)/1--0 S(1) is lower for thermal excitation 
than for UV
fluorescence, while {\iron}/{\brg} is higher in X-ray or shock excited
regions than in {\hii} regions (e.g. Colina 1993).  X-ray excitation
can also produce lower \iron/{\brg} ratios, but this would lead to
extreme 1--0 S(1)/{\brg} $>$ 100. The H$_2$ line ratios 2--1 S(1)/1--0
S(1) and 1--0 S(0)/1--0 S(2) can be used to derive the vibrational
excitation temperature $T_{\rmn{vib}}$ and the rotational temperature
$T_{\rmn{rot}}$, respectively.

The mass of the excited molecular hydrogen in the nucleus was derived
using the transition probabilities by Wolniewicz, Simbotin \& Dalgarno
(1998), the vibration temperature and thermal equilibrium, and are
given in Table \ref{prm}. If $T_{\rmn{vib}}$ is unknown,
it is assumed to be 2000 K. This leads to a factor of $\sim$2
uncertainty in the mass, if the real temperature is $>$1500
K. Below 1500 K the fraction of H$_2$ in the $v=1,J=3$ energy level
declines rapidly and the mass is underestimated by a factor of $\sim$10 if
$T=1000$ K. However, 1--0 S(1) emission is likely to be biased towards
warmer gas, because the $E_{\rmn{upper}} = 6950$ K for the $v=1,J=3$
level. We have also determined the average surface density of the
excited {\h2} in the nucleus.

The Gaussian FWHM sizes for H$_2$ and {\iron} in Table \ref{prm} have
been obtained by measuring the line flux in each column across the
nuclear region and are corrected for the seeing.  The number of
columns fitted was chosen to include the nuclear knot and exclude any
additional extended emission. The structures probed here correspond to
physical scales of 100--200 pc, depending on the distance of the
galaxies.

\begin{table*}
  \begin{minipage}{170mm}
    \caption{Dereddened emission line ratios relative to Br$\gamma$.
      The galaxies from Paper I are included to facilitate comparison.
      For NGC 1386 a 0.6 arcsec slit was used.\label{ratio}}
    \begin{tabular}{lllllllllllllll}
      \hline
      
                 &1068   &1097   &1365    &1386   &1566    &2110    &E428   &2992    &3081   &3227    &3783    &4945   &5128   &7582   \\
      Nucleus    &Sy2    &Sy1    &Sy1.8   &Sy2    &Sy1     &Sy2     &Sy2    &Sy2     &Sy2    &Sy1.5   &Sy1     &Sy2    &Sy2    &Sy2  \\\\
      \iron      &4.0    &$>$2.2 &0.36    &6.84   &0.48    &8.34    &2.55   &0.93    &0.64   &1.47    &0.22    &3.77   &14.6   &0.53   \\
      \sivi      &13.8   &...    &...     &6.03   &...     &...     &2.66   &0.39    &8.73   &0.36    &0.42    &...    &...    &0.37   \\
      1--0 S(2)  &$<$0.3 &$>$4.3 &$<$0.04 &0.81   &0.05    &0.56    &0.23   &0.09    &0.91   &0.24    &$<$0.02 &1.49   &$<$0.2 &0.11   \\
      \he        &$<$0.3 &...    &0.41    &$<$0.2 &$<$0.04 &$<$0.07 &0.28   &0.39    &0.37   &$<$0.02 &0.35    &0.35   &$<$0.2 &$<$0.03\\
      1--0 S(1)  &0.42   &$>$4.4 &$<$0.04 &1.14   &0.28    &0.93    &0.84   &0.28    &1.60   &0.54    &0.21    &3.10   &2.03   &0.07   \\
      1--0 S(0)  &$<$0.3 &$>$1.4 &$<$0.04 &$<$0.2 &$<$0.04 &0.30    &0.18   &0.06    &0.21   &0.15    &$<$0.02 &0.53   &0.34   &$<$0.02\\
      2--1 S(1)  &$<$0.2 &...    &$<$0.04 &$<$0.2 &$<$0.04 &0.24    &$<$0.2 &0.05    &$<$0.2 &0.04    &$<$0.02 &0.35   &$<$0.2 &$<$0.03\\
      \caviii    &1.46   &...    &$<$0.05 &2.09   &$<$0.06 &$<$0.1  &$<$0.2 &$<$0.06 &1.66   &$<$0.01 &0.08    &$<$0.1 &$<$0.2 &$<$0.03\\
      \sivii     &7.96   &...    &$<$0.12 &9.58   &$<$0.3  &$<$0.8  &1.7    &$<$2.0  &7.33   &0.24    &0.46    &$<$0.2 &$<$0.3 &0.45   \\
      \hline
    \end{tabular}
  \end{minipage}
\end{table*}

\section{Results}

\begin{figure*}
  \begin{minipage}[t]{0.33\linewidth}
    \includegraphics[height=5.85cm]{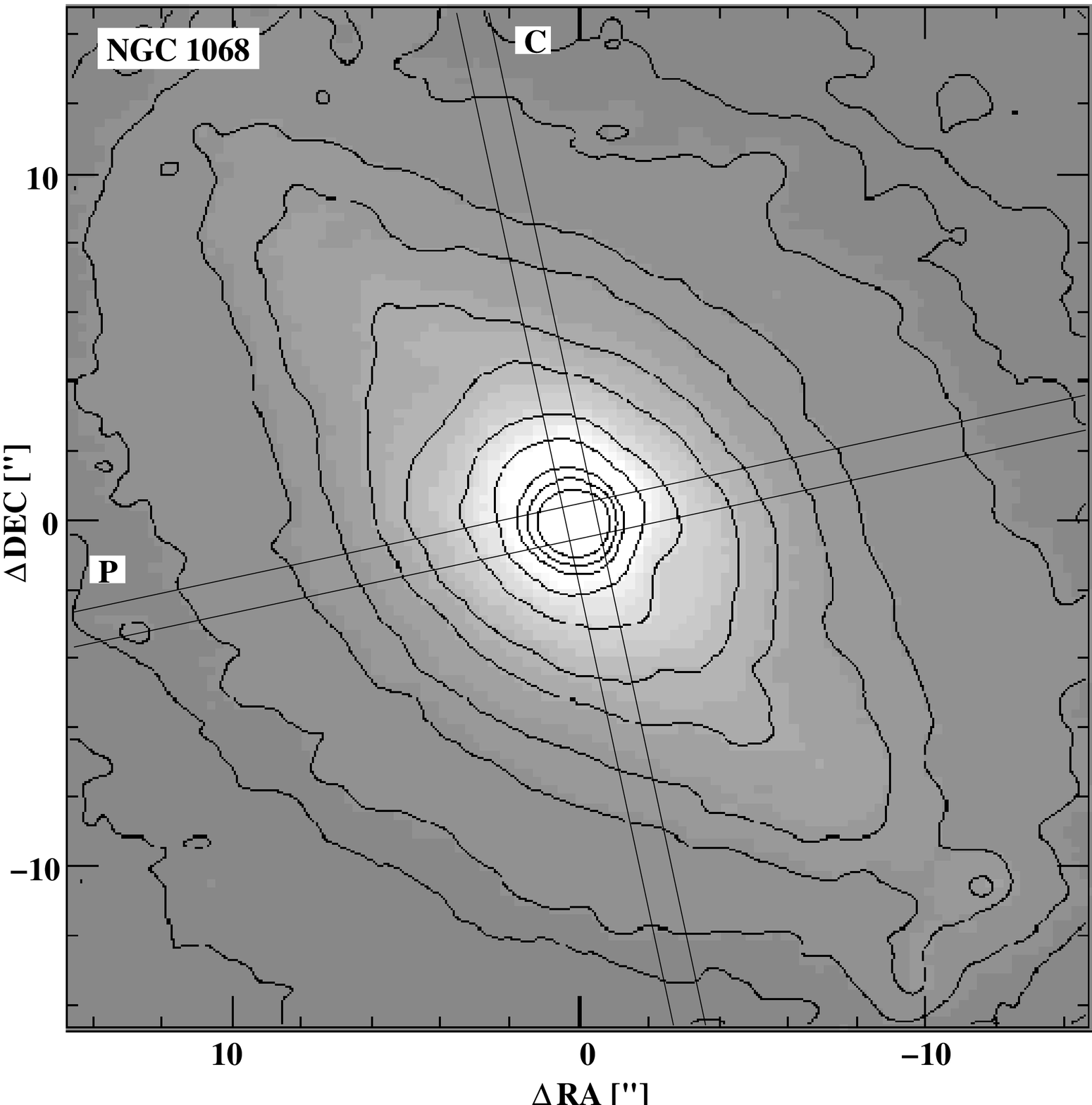}
  \end{minipage}
  \begin{minipage}[t]{0.33\linewidth}
    \includegraphics[height=5.85cm]{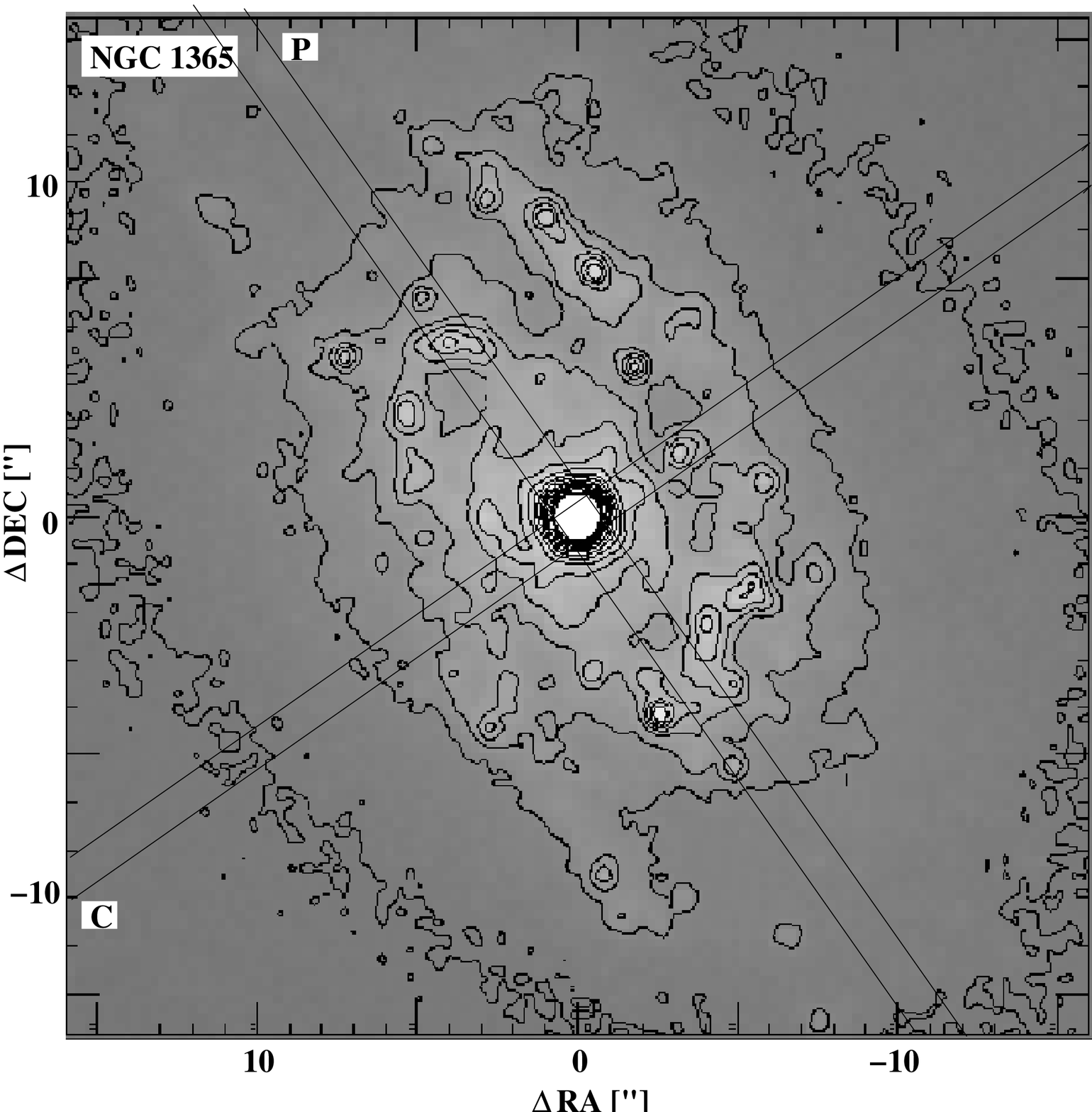}
  \end{minipage}
  \begin{minipage}[t]{0.33\linewidth}
    \includegraphics[height=5.85cm]{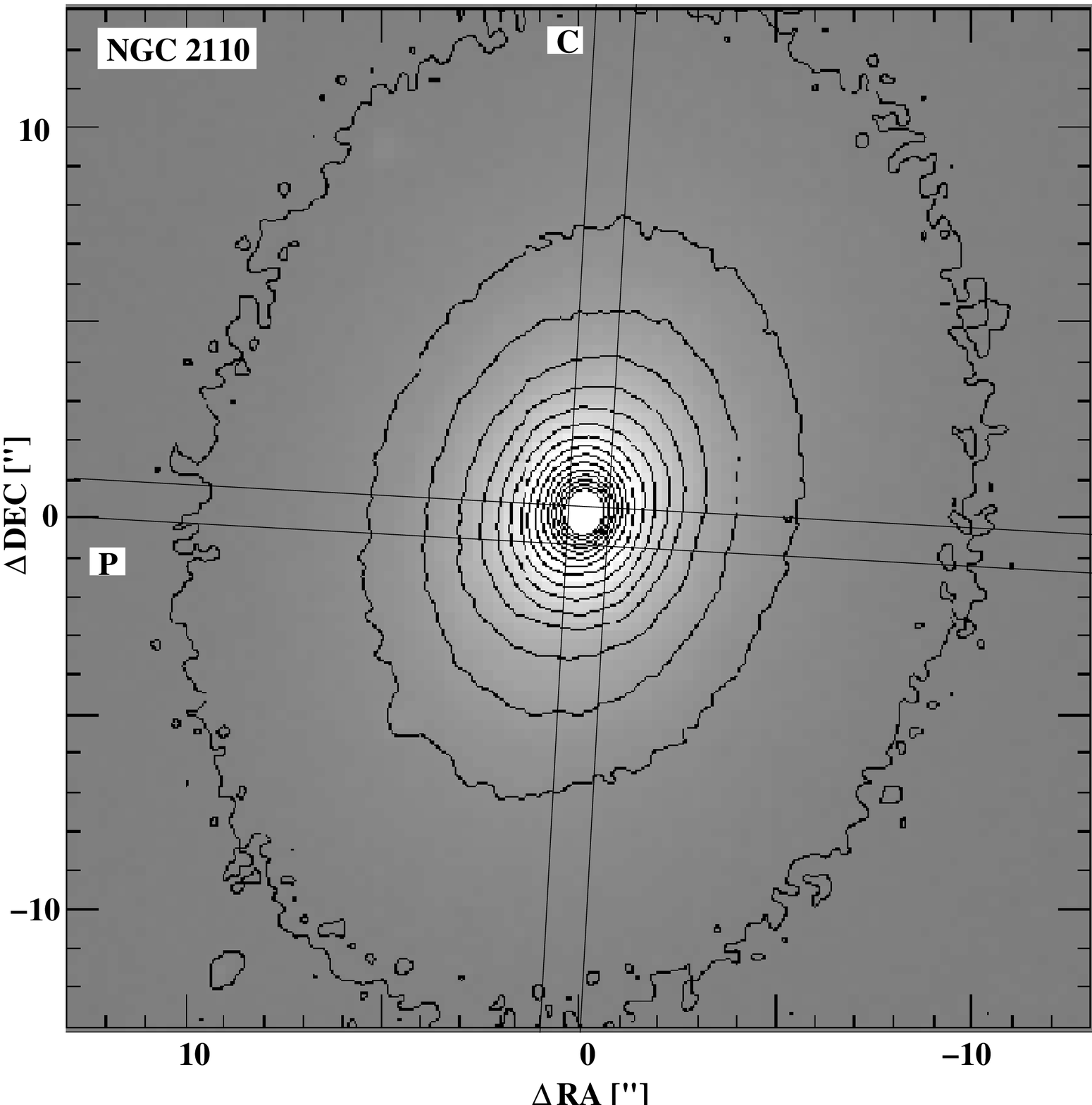}
  \end{minipage}

  \begin{minipage}[t]{0.33\linewidth}
    \includegraphics[height=5.85cm]{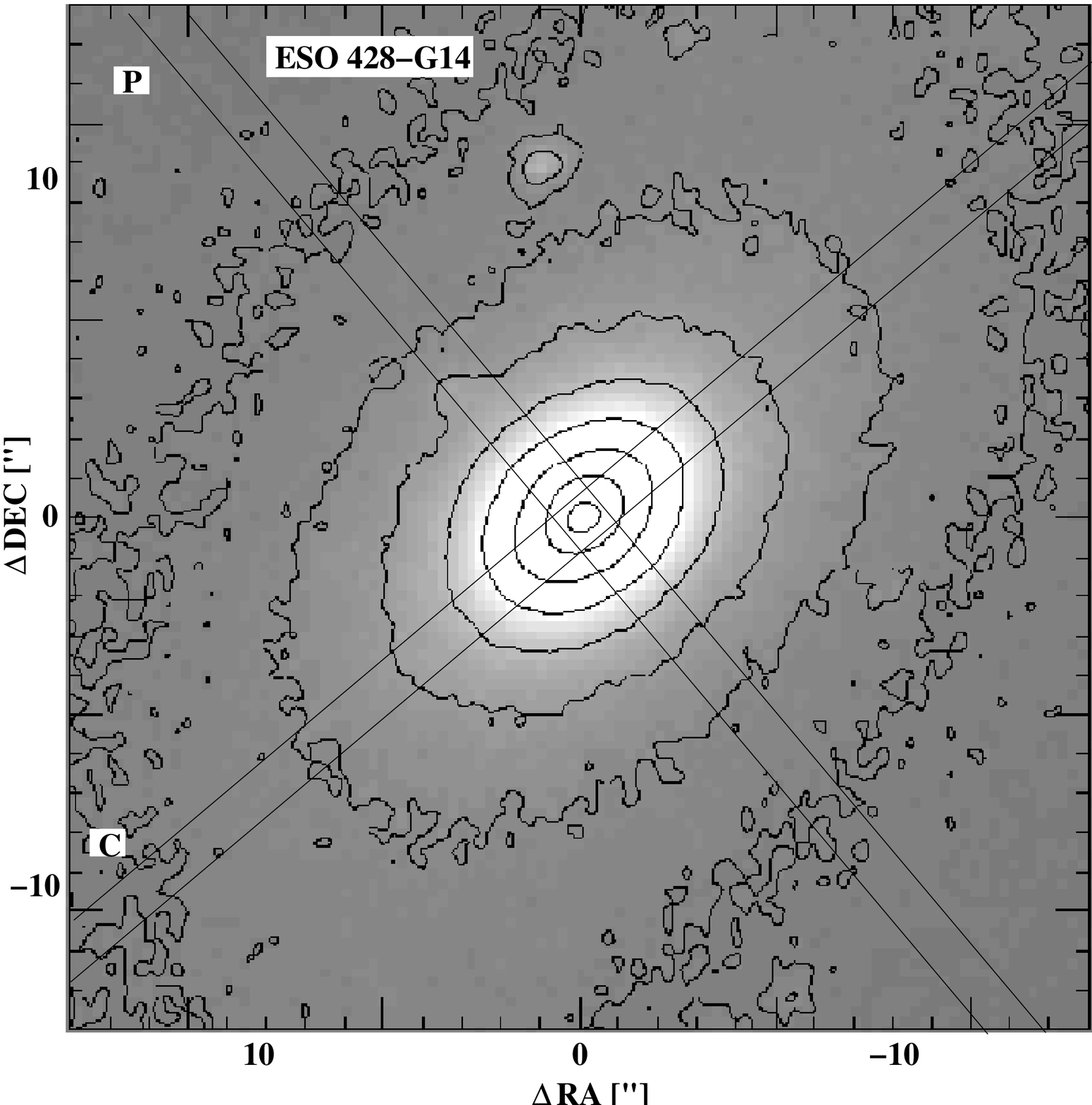}
  \end{minipage}
  \begin{minipage}[t]{0.33\linewidth}
    \includegraphics[height=5.85cm]{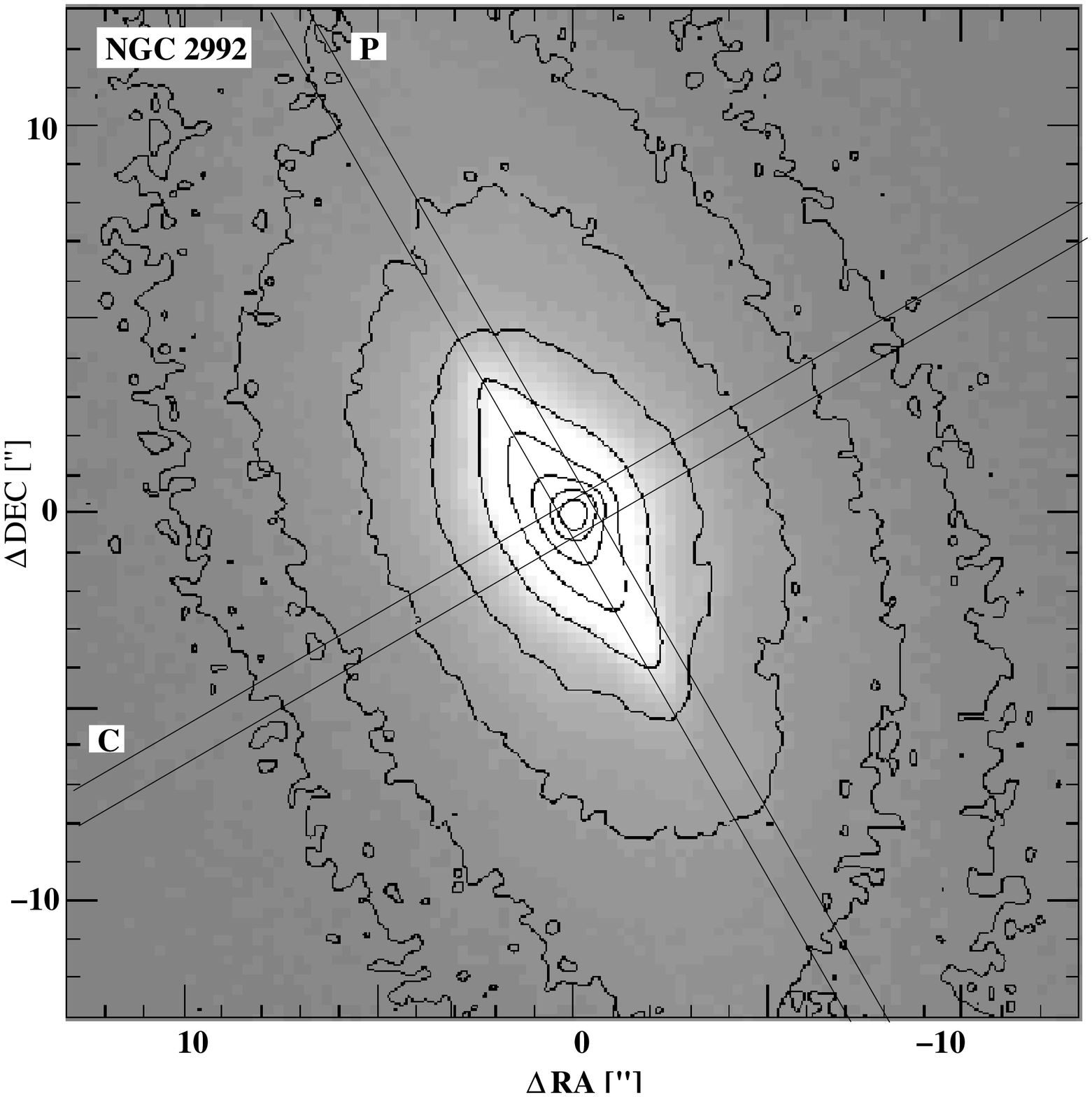}
  \end{minipage}
  \begin{minipage}[t]{0.33\linewidth}
    \includegraphics[height=5.85cm]{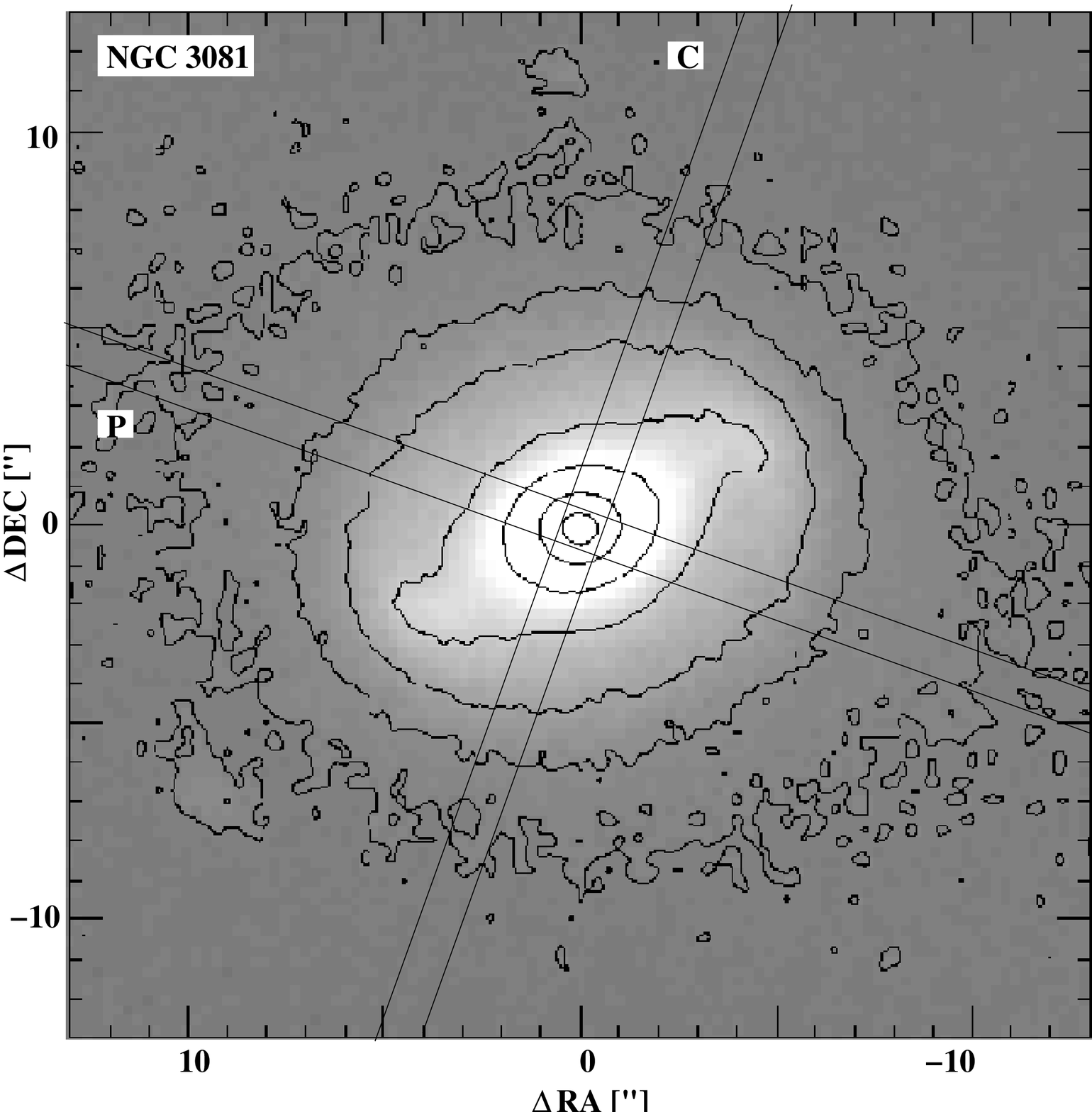}
  \end{minipage}

  \begin{minipage}[t]{0.33\linewidth}
    \includegraphics[height=5.85cm]{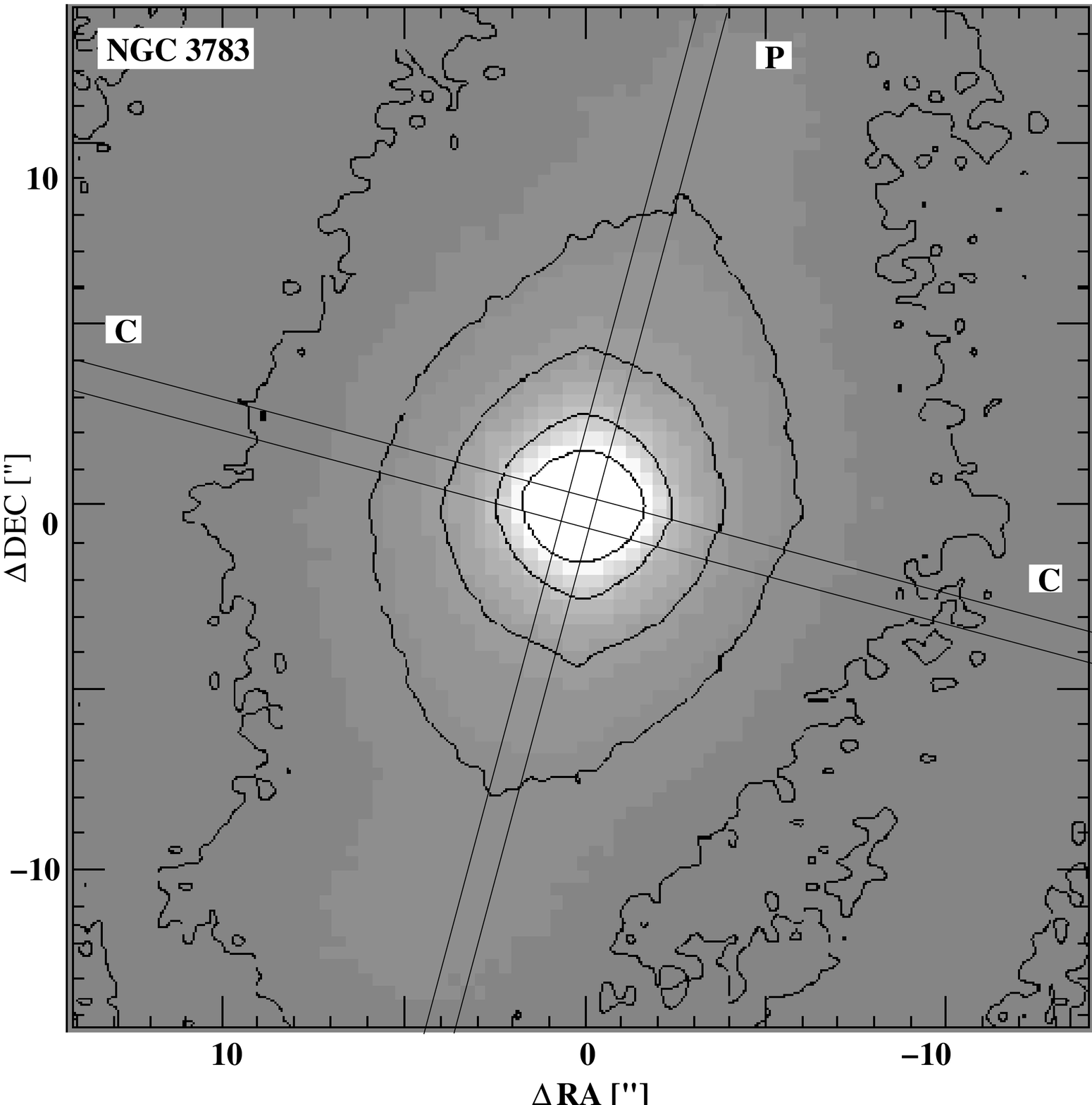}
  \end{minipage}
  \begin{minipage}[t]{0.33\linewidth}
    \includegraphics[height=5.85cm]{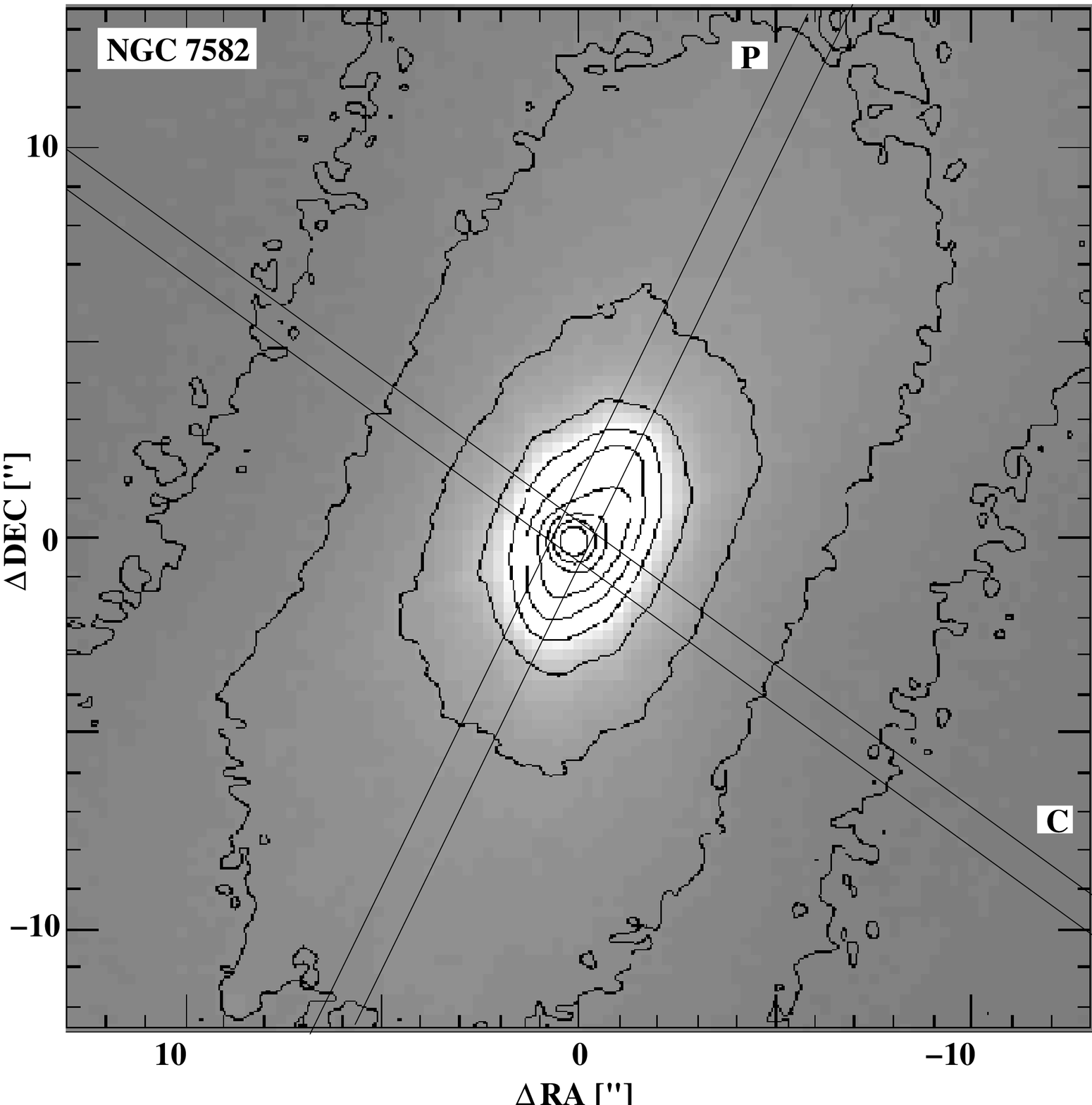}
  \end{minipage}
  \caption{The near-infrared images of NGC 1068 ({\em top left}; $H$-band from 
    Peletier et al. 1999), NGC 1365 ({\em top
      center}), NGC 2110 ({\em top right}), ESO 428-G14 ({\em middle
      left}), NGC 2992 ({\em middle center}), NGC 3081 ({\em
      middle right}, NGC 3783 ({\em bottom
      left}; $K$-band from Mulchaey et al. (1997)) and 
    NGC 7582 ({\em bottom right}). All images are in the $Ks$-band except as 
    marked. North is up and East to the left.
    Slit positions are indicated in the images, marked with C for
    parallel to the cone, on the side where the cone is stronger, and
    P perpendicular to the cone.
    \label{broad}}
\end{figure*}

\begin{figure*}
  \includegraphics[height=20.1cm]{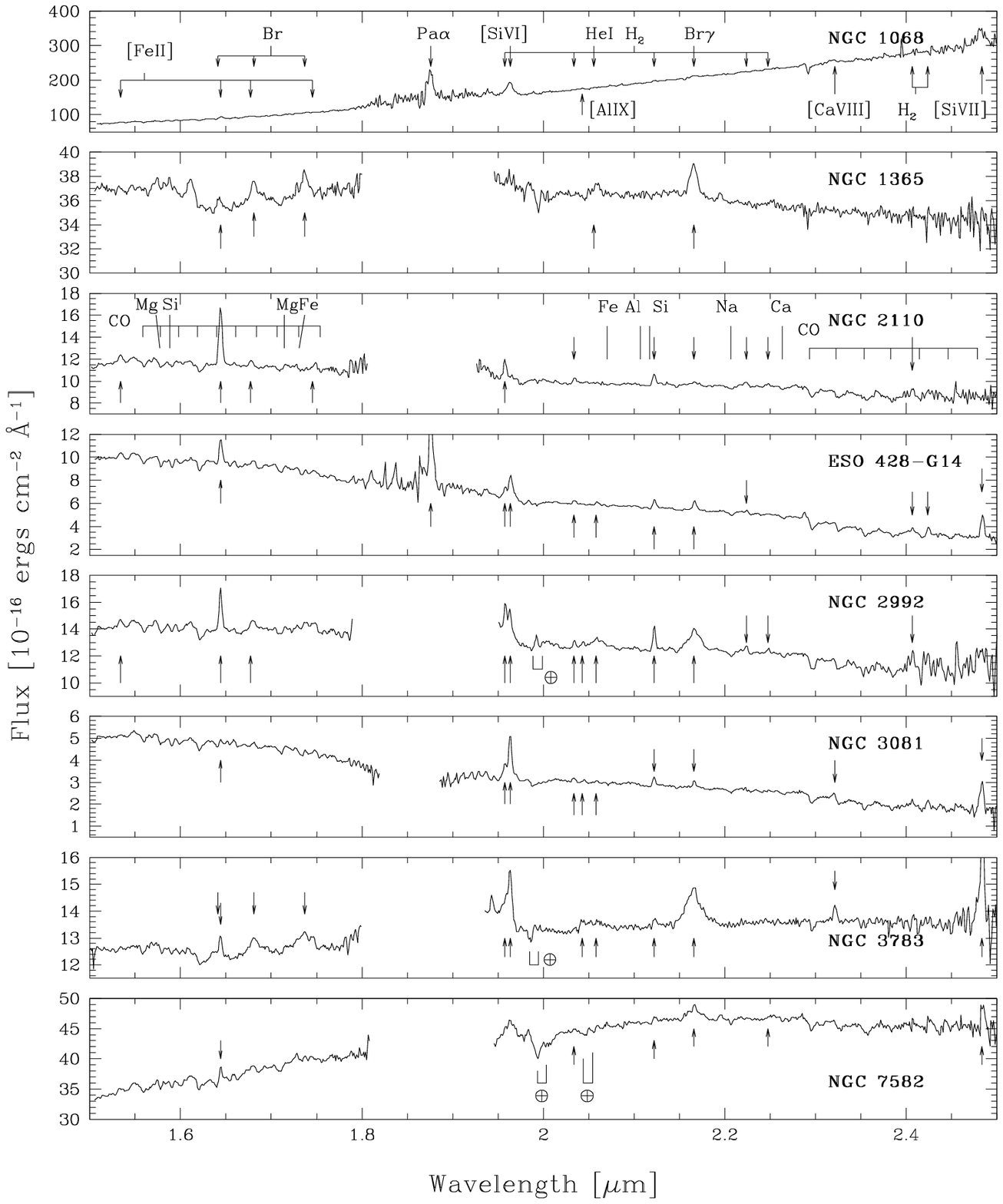}
  \caption{The nuclear $H$- and $K$-band spectra in a nuclear 1.4 arcsec aperture for, 
    from top to bottom, NGC 1068, NGC 1365, NGC 2110, ESO 428-G14, NGC
    2992, NGC 3081, NGC 3783 and NGC 7582. All the detected emission lines
    have been labelled in the
    first panel, and marked with arrows in the subsequent panels. The main
    absorption lines have been marked in the panel of NGC 2110.
    Atmospheric features are marked with $\oplus$.
    \label{allhk}}
\end{figure*}

\begin{figure*}
  \begin{minipage}[t]{0.4\linewidth}
    \includegraphics[width=7cm]{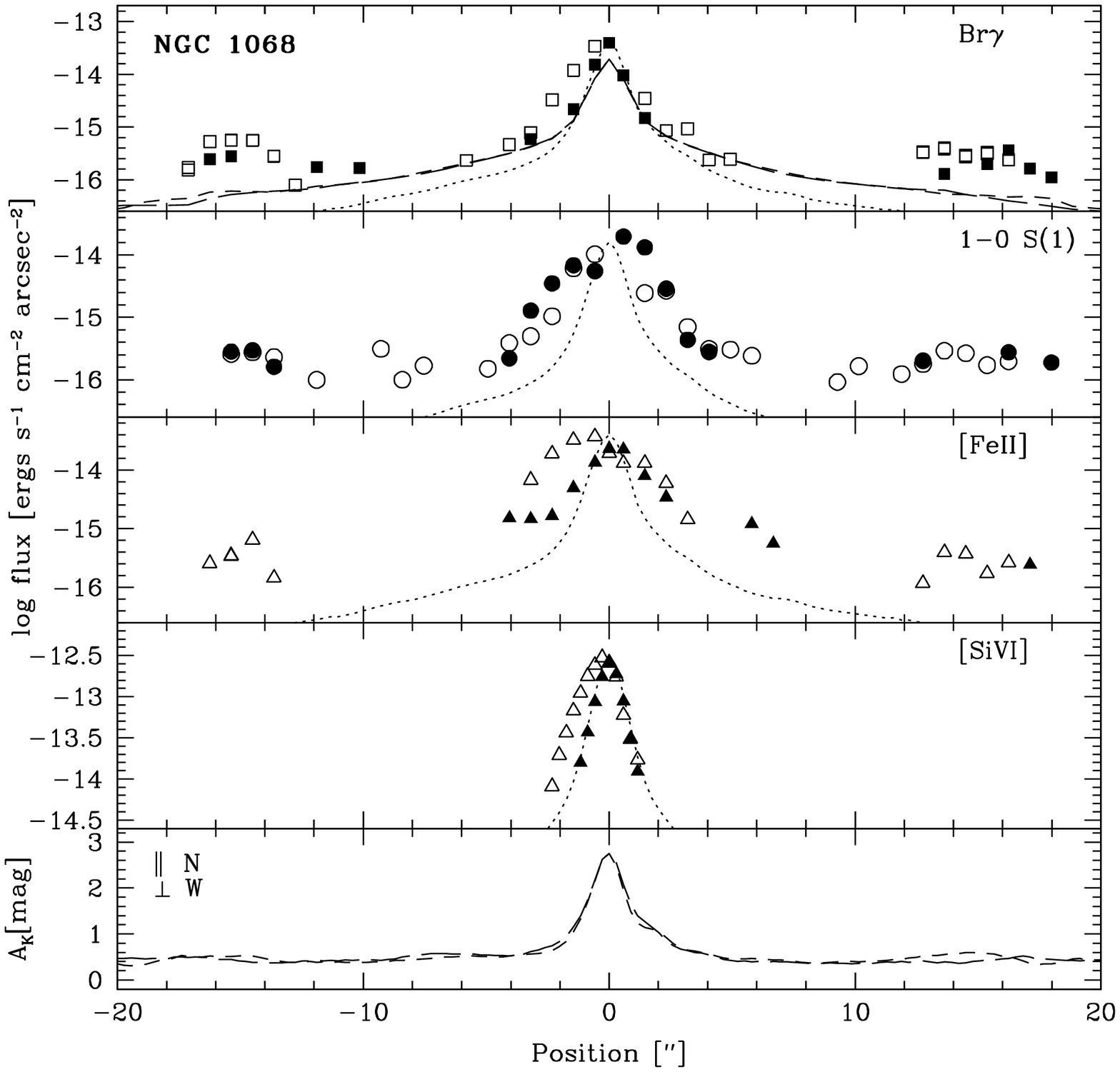}
  \end{minipage}
  \begin{minipage}[t]{0.4\linewidth}
    \includegraphics[height=7cm]{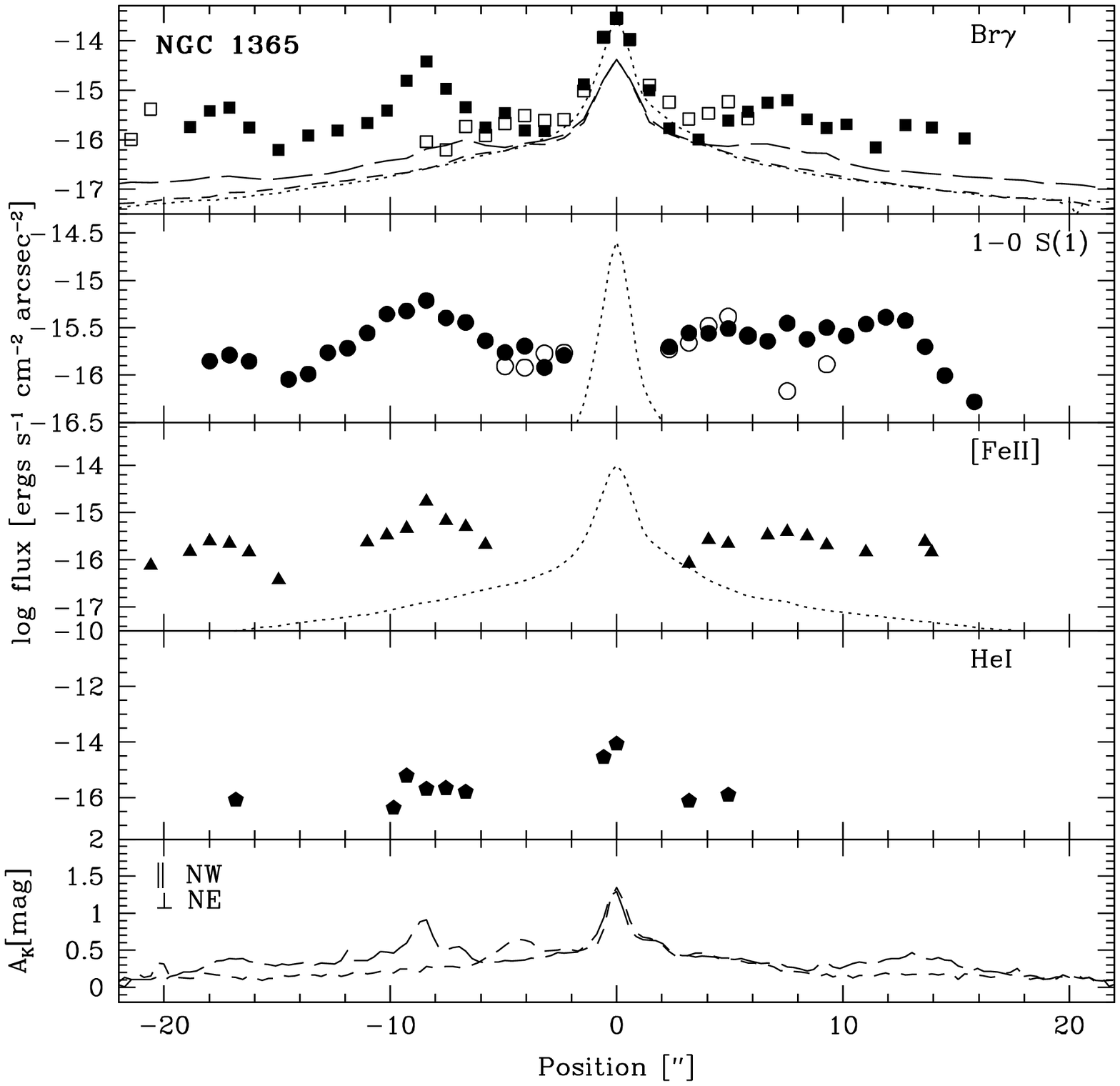}
  \end{minipage}

  \begin{minipage}[t]{0.4\linewidth}
    \includegraphics[height=7cm]{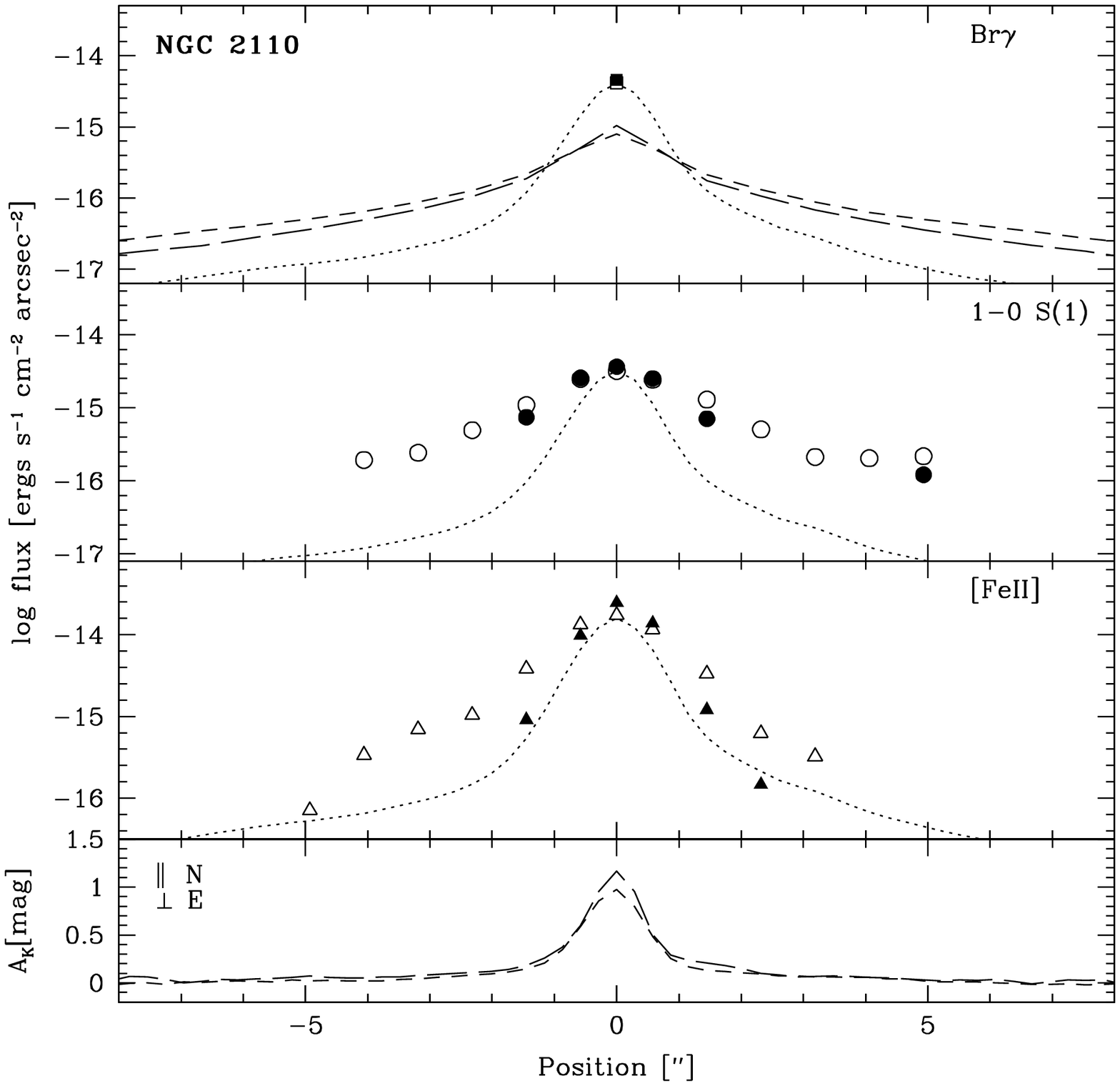}
  \end{minipage}
  \begin{minipage}[t]{0.4\linewidth}
    \includegraphics[height=7cm]{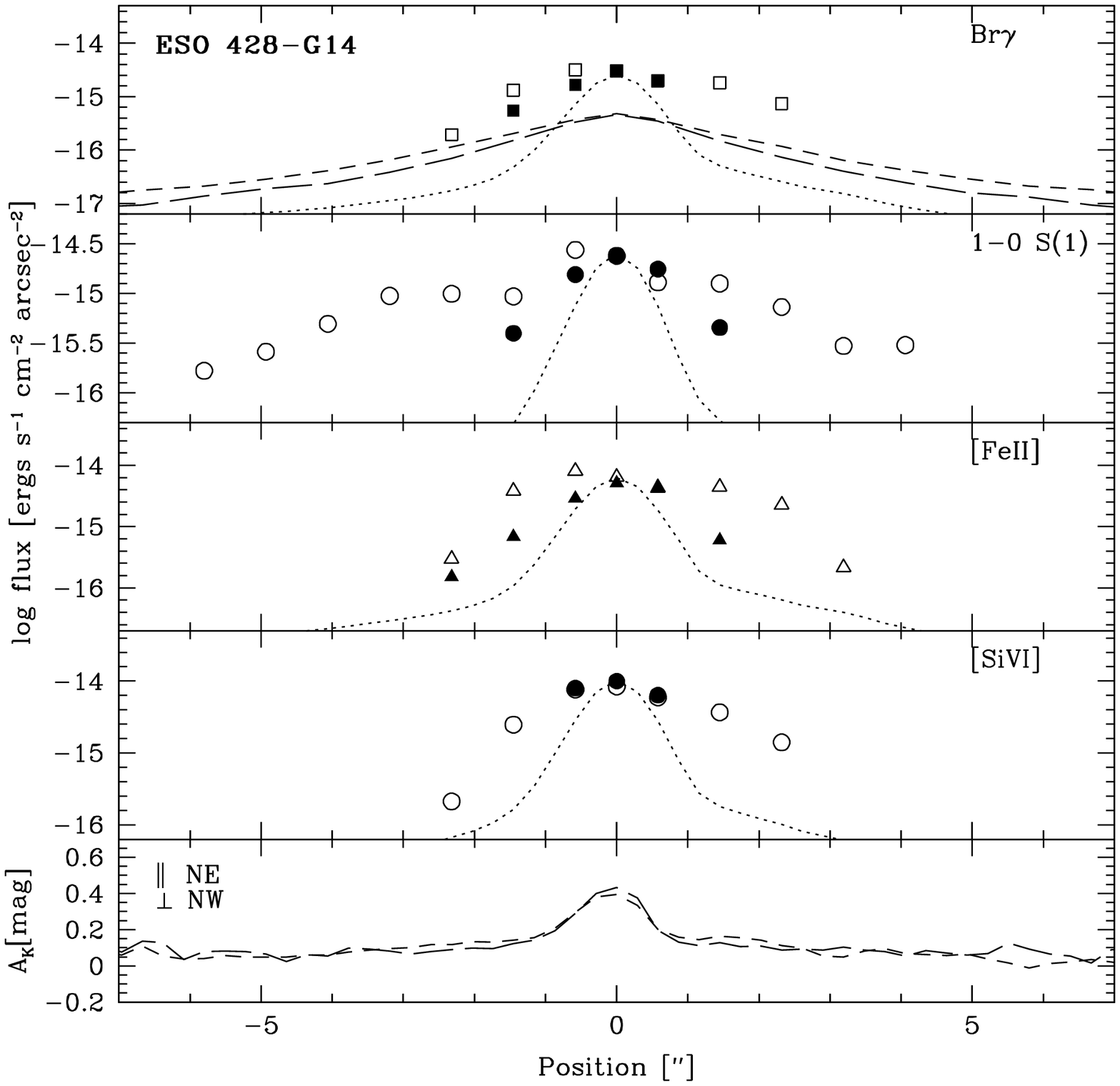}
  \end{minipage}

  \begin{minipage}[t]{0.4\linewidth}
    \includegraphics[height=7cm]{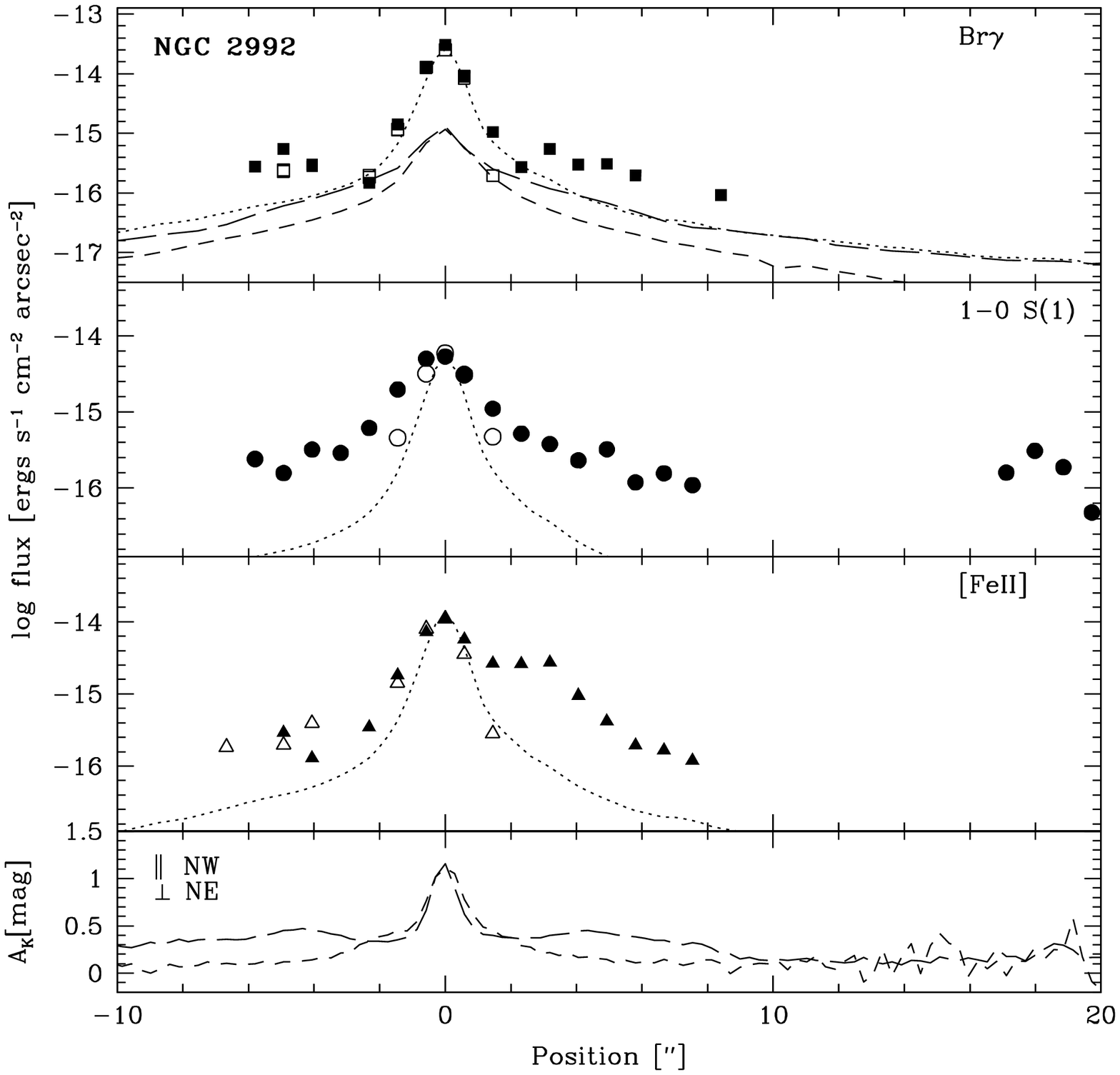}
  \end{minipage}
  \begin{minipage}[t]{0.4\linewidth}
    \includegraphics[height=7cm]{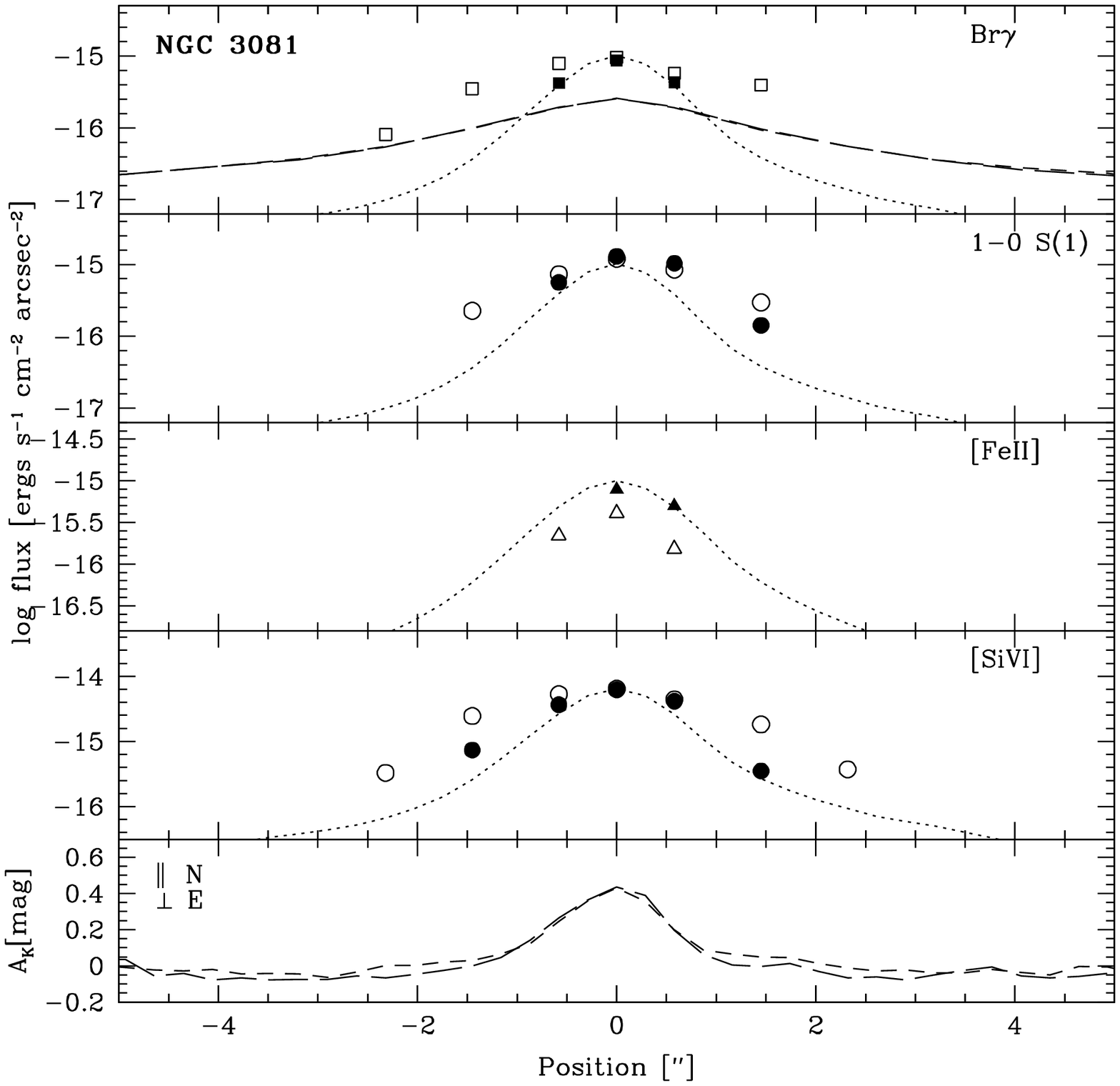}
  \end{minipage}

  \caption{The observed line emission in NGC 1068 ({\em upper left}), 
    NGC 1365 ({\em upper right}), NGC 2110 ({\em middle left}), ESO
    428-G14 ({\em middle right}), NGC 2992 ({\em lower left}) and NGC
    3081 ({\em lower right}). Open symbols indicate parallel to the
    cone and filled symbols perpendicular to the cone. The PSF is
    indicated by dotted lines.  Extinction is plotted in the lowermost
    panel of each sub-figure. Negative direction along the slit has
    been marked. In the uppermost panel, the 2.1 $\mu$m continuum
    emission is indicated parallel to the cone ({\em short-dashed
      line}) and perpendicular to it ({\em long-dashed line}). In most
    cases three columns have been binned together to lower noise.
    \label{emission}}
\end{figure*}

\begin{figure*}
  \includegraphics[]{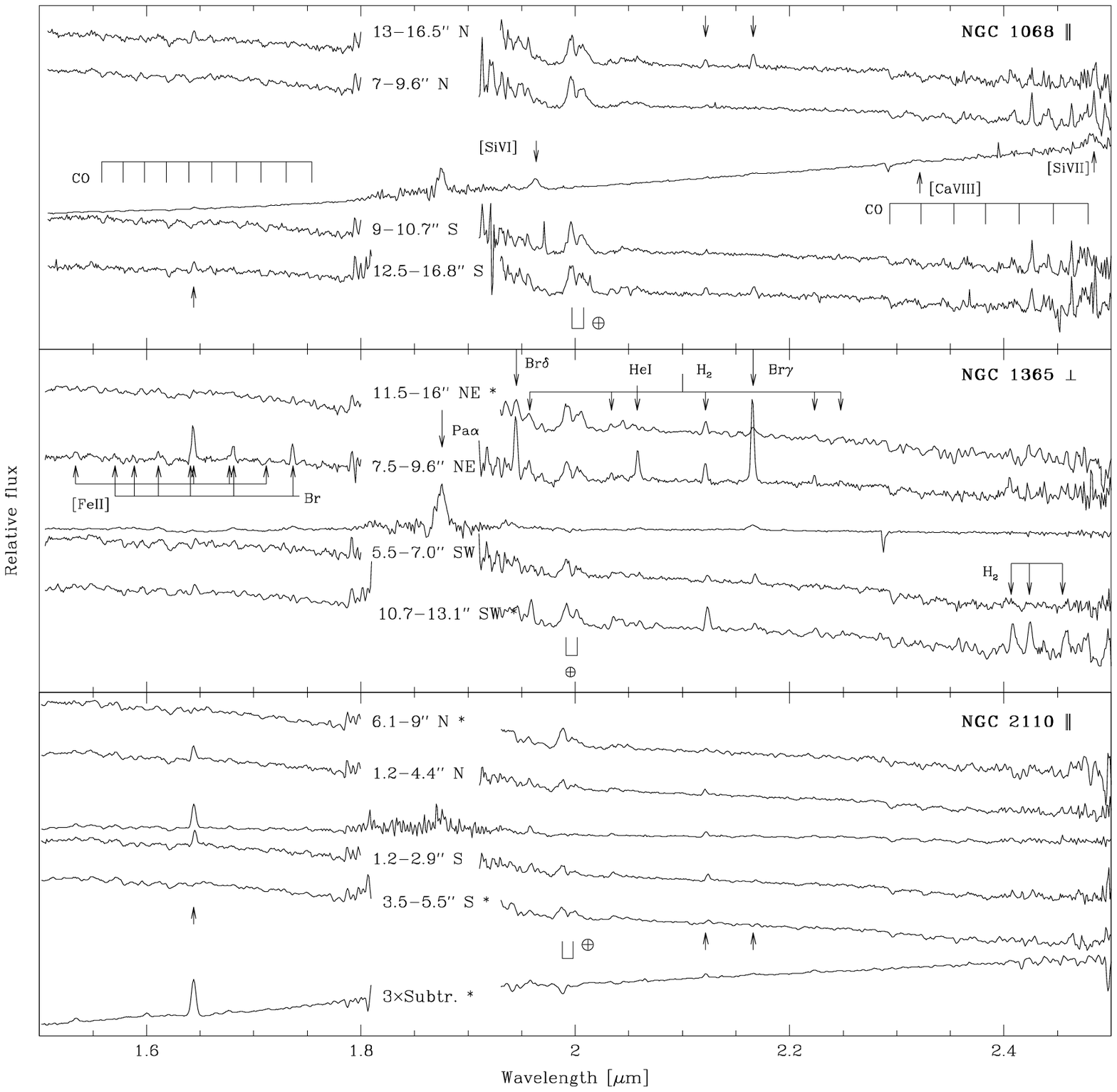}
  \caption{The 1.5--2.5 $\mu$m spectra of, from top to bottom, NGC 1068, 
    NGC 1365 and NGC 2110, at different positions along the slit. The
    K-band flux was scaled to match the flux in the nuclear 1.4 arcsec
    aperture, and the spectra were placed at uniform intervals to
    facilitate comparison between different regions. The absorption
    features at 2.0 $\mu$m and 2.05 $\mu$m are due to imperfect atmospheric
    cancellation. The nuclear subtracted spectrum was obtained as
    discussed in the text. The spectra marked with a star have been
    smoothed with three pixel boxcar function. The direction with
    respect to the cone has been marked next to the label on the upper
    right corner in each panel.
    \label{sp1}}
\end{figure*}

\begin{figure*}
  \begin{minipage}[t]{0.4\linewidth}
    \includegraphics[height=7cm]{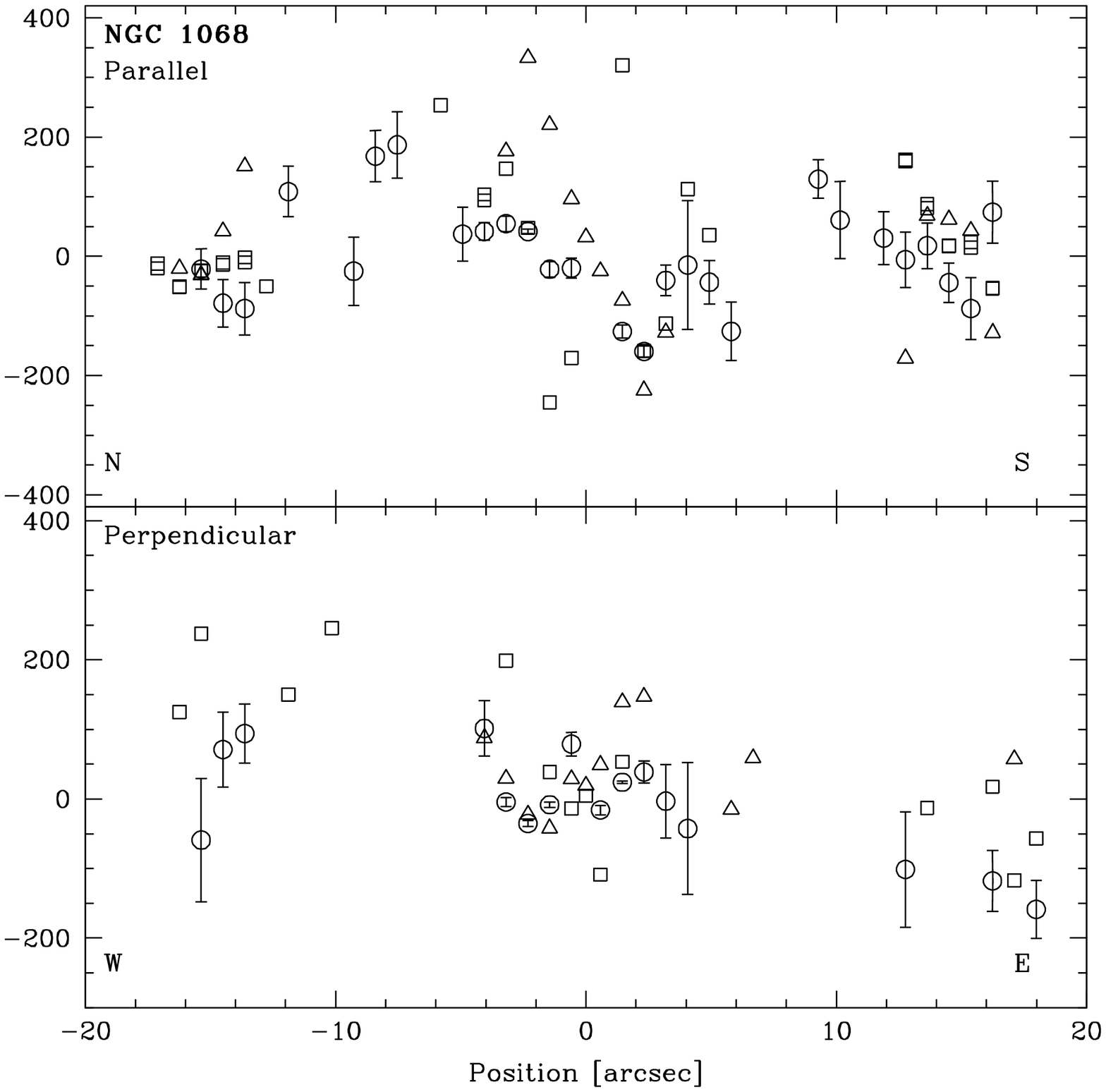}
  \end{minipage}
  \begin{minipage}[t]{0.4\linewidth}
    \includegraphics[height=7cm]{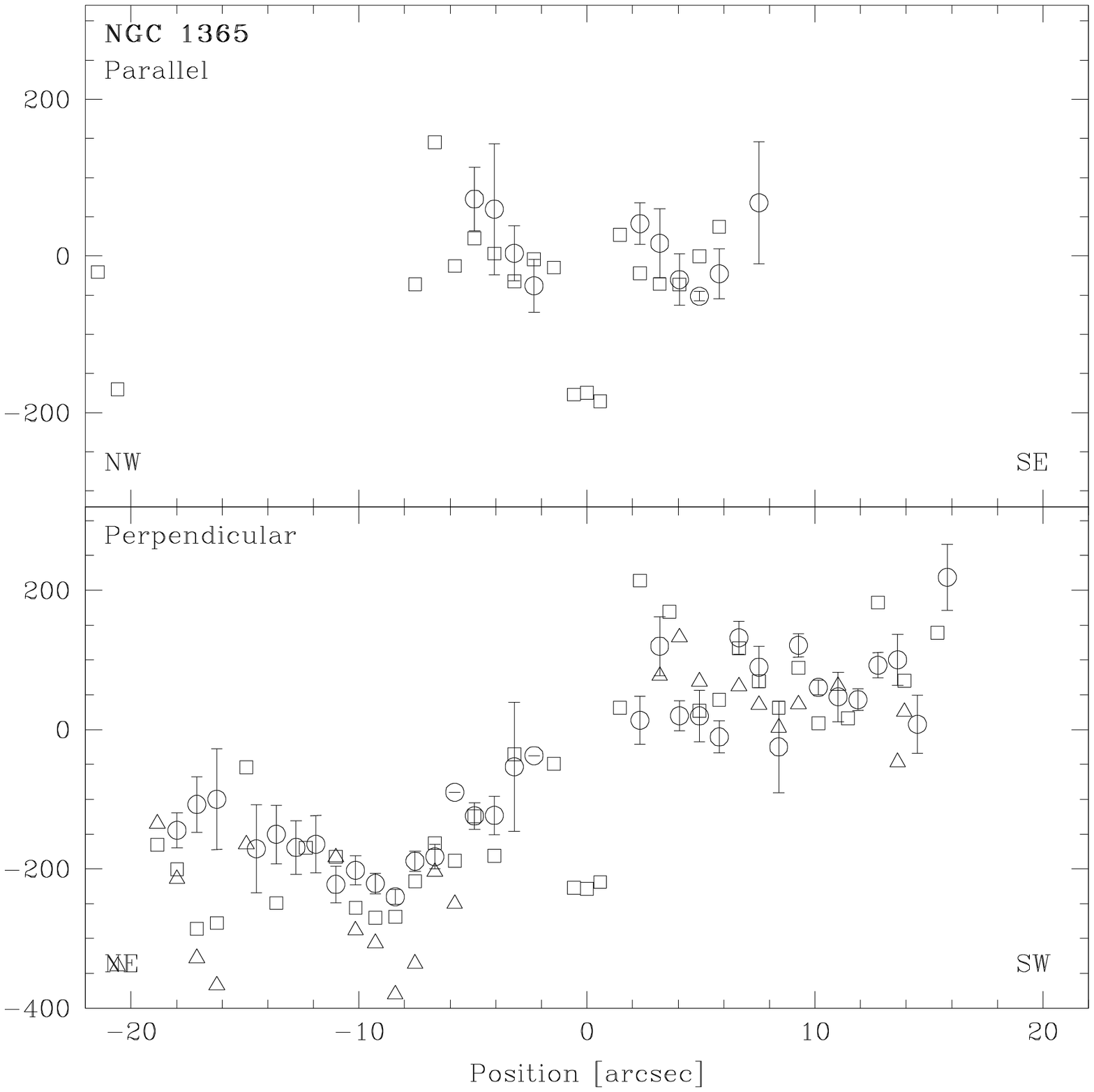}
  \end{minipage}

  \begin{minipage}[t]{0.4\linewidth}
    \includegraphics[height=7cm]{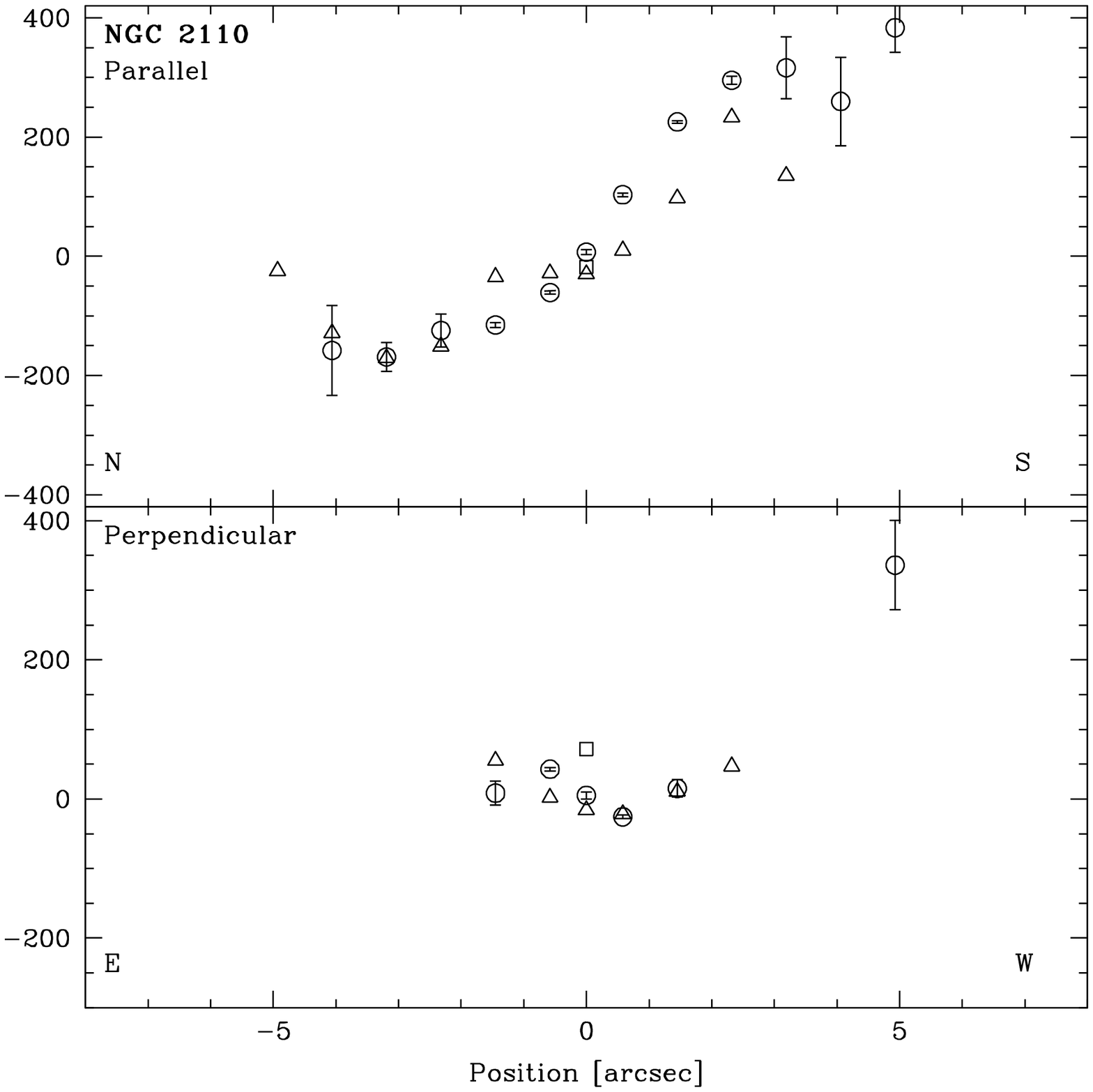}
  \end{minipage}
  \begin{minipage}[t]{0.4\linewidth}
    \includegraphics[height=7cm]{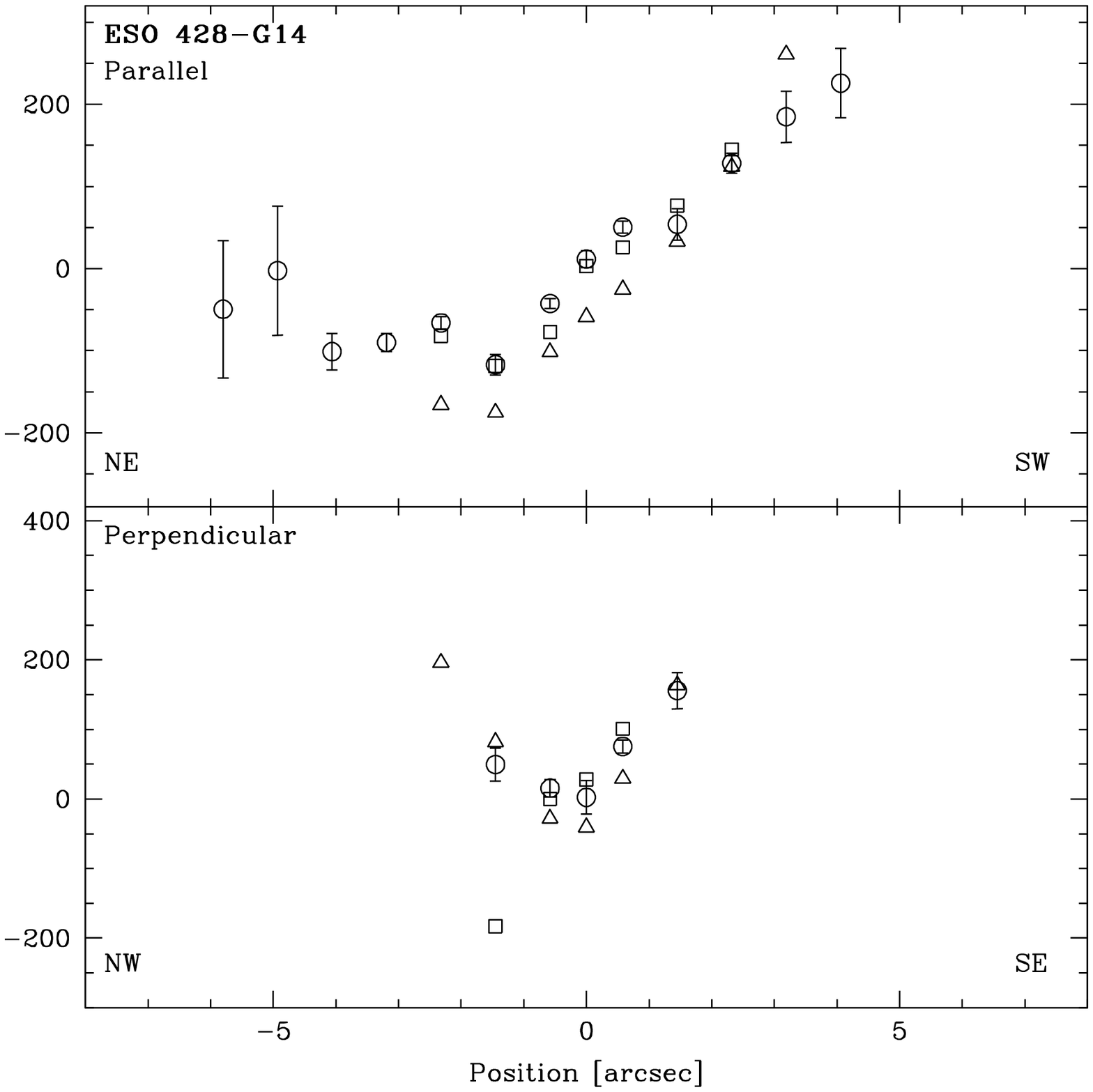}
  \end{minipage}

  \begin{minipage}[t]{0.4\linewidth}
    \includegraphics[height=7cm]{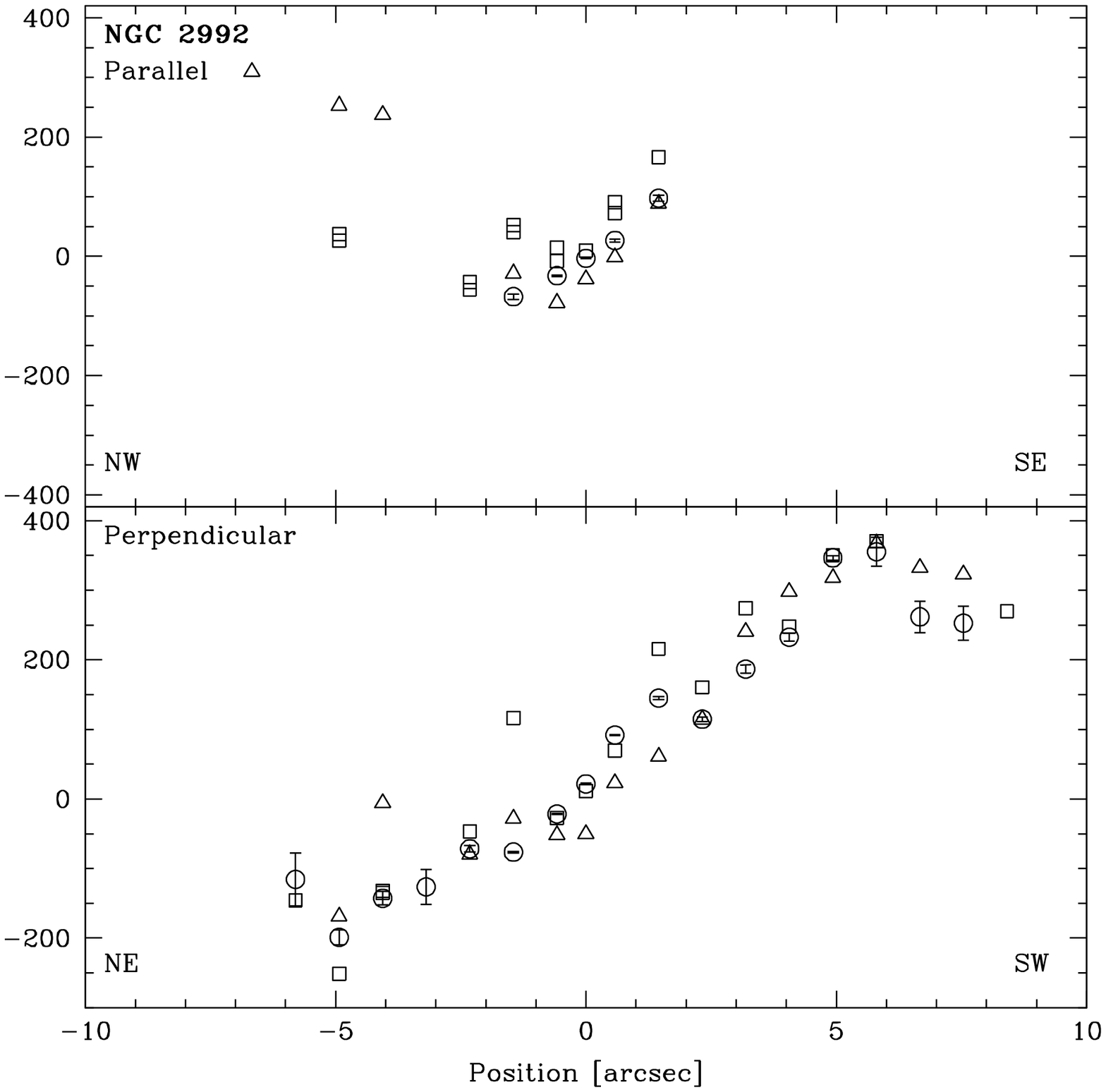}
  \end{minipage}
  \begin{minipage}[t]{0.4\linewidth}
    \includegraphics[height=7cm]{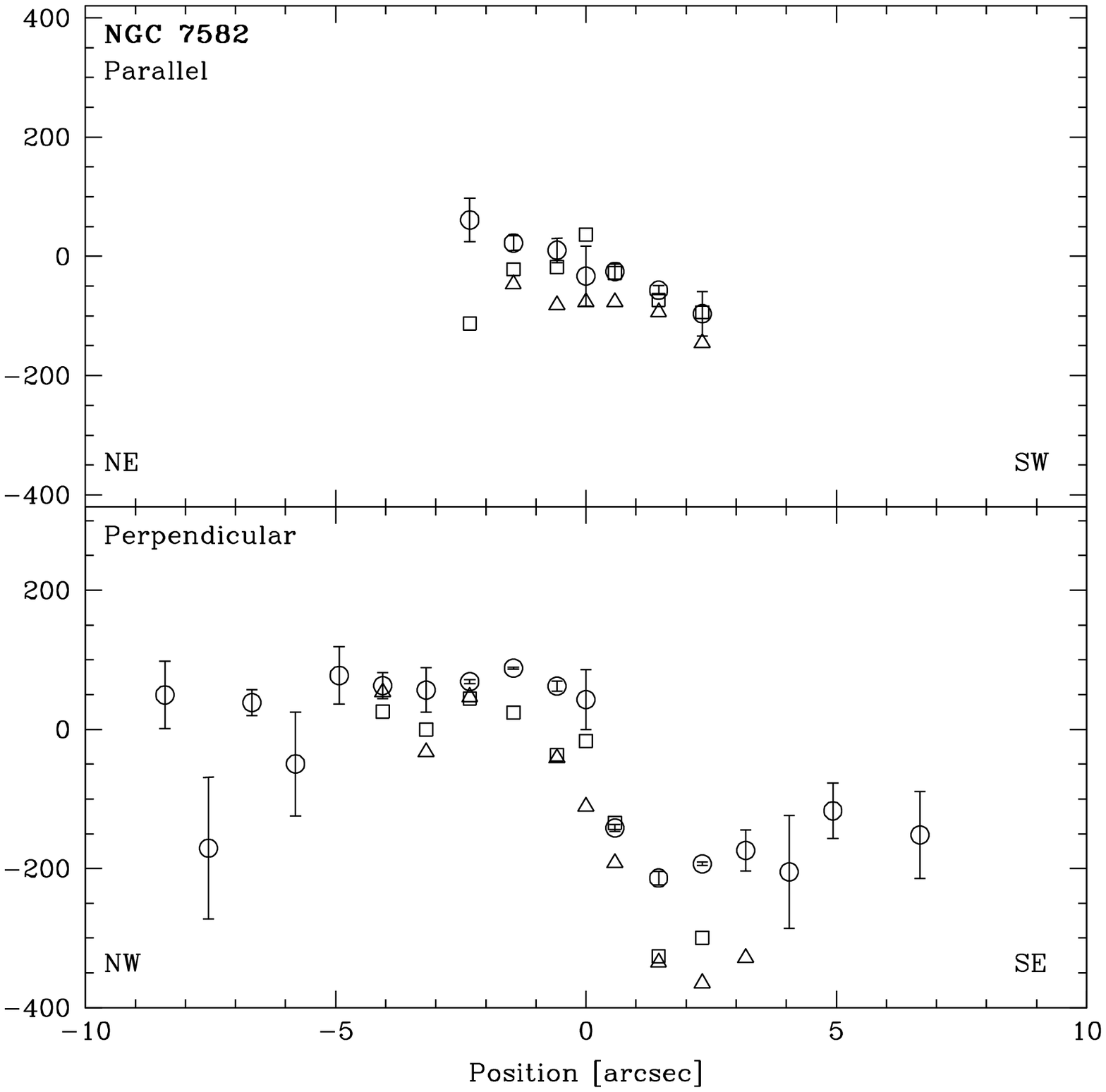}
  \end{minipage}
  \caption{The observed velocity field in units of {\kms} 
    in the 1--0 S(1) ({\em circles} with error bars), Br$\gamma$ ({\em
      squares}) and {\iron} ({\em triangles}) lines for NGC 1068({\em
      upper left}), NGC 1365 ({\em upper right}), NGC 2110 ({\em
      middle left}), ESO 428-G14 ({\em middle right}), NGC 2992 ({\em
      bottom left}) and NGC 7582 ({\em bottom right}). The x-axis
    shows the position along the slit.
    \label{velocity}}
\end{figure*}

The nuclear 30$\times$30 arcsec $K$-band images of the galaxies, with
the slit positions overlaid, are shown in Fig. \ref{broad} and the
$H$- and $K$-band nuclear spectra in Fig. \ref{allhk}.  The spatial
extent of the main emission lines is shown in Figs. \ref{emission} and
\ref{emission2}, velocity fields of the main emission lines in Fig.
\ref{velocity}, and the spectra at different distances from the
nucleus in Figs. \ref{sp1}, \ref{sp2} and \ref{sp3}.  The
nuclear-subtracted spectra have been obtained by matching the flux of
the $K$-band CO lines in the off-nuclear (0.7-1.6 arcsec radius from
the nucleus) spectra to the 1.4 arcsec diameter nuclear aperture and
subtracting the result from the nuclear spectra. The
nuclear-subtracted spectra are included here to better show the weak
nuclear emission lines. The subtracted spectra are not shown for NGC
1068, NGC 1365 or NGC 3783 because these galaxies have very weak CO
lines (CO index $\sim$0).

\subsection{NGC 1068\label{section1068}}

NGC 1068 is a nearby ($z$ = 0.00379; distance = 15.1 Mpc) (R)SA(rs)b
galaxy. It has an extensively studied Sy2 nucleus with high extinction
derived towards the BLR (A$_V$ $>$50 mag), as evidenced by the
non-detection of broad Br$\alpha$ 4.05 $\mu$m (Lutz et al. 2002). NGC
1068 also harbors a star forming ring with diameter $\sim$15 arcsec
($\sim$1.1 kpc; e.g. Davies, Sugai \& Ward 1998).  Rouan et al. (1998)
found K-band extended emission along PA = 102\degr, up to 15 pc
radius, interpreted as the equatorial plane of an inclined torus. This
extension can also be seen in the $L$$M$-bands (Marco \& Alloin
2000). NGC 1068 has an {\oiii} cone, which extends to $\sim$7.5 arcsec
at PA = 35\degr (Evans et al. 1991).

The $H$- and $K$-band nuclear spectrum of NGC 1068 (Fig. \ref{allhk})
resembles that of NGC 5128 (Paper I), having an almost power-law like
continuum shape. The most prominent features in the nuclear spectra
are the coronal lines {\sivi} 1.964 $\mu$m, {\caviii} 2.321 $\mu$m and
{\sivii} 2.483 $\mu$m. The other emission lines detected in the
nucleus are {\iron} 1.644 $\mu$m, \h2 1--0 S(1) 2.122 $\mu$m and
{\brg} 2.166 $\mu$m.  Also Pa$\alpha$ 1.876 $\mu$m is prominent, but
its usefulness is limited by telluric residuals. Nuclear H$_2$
emission is faint.

1--0 S(1) is the most extended line and is detected up to 5--7 arcsec
(300 pc) from the nucleus parallel to the cone and $\sim$4 arcsec
perpendicular to it.  {\iron} and {\brg} are extended $\sim$4 arcsec
from the nucleus.  \brg, \iron, \h2 and {\he} 2.058 $\mu$m are
furthermore detected in the star forming ring $\sim$15 arcsec
($\sim$1.1 kpc) from the nucleus.  {\sivi} 1.964 $\mu$m is the
strongest emission line in the nucleus, {\iron} between the nucleus
and the ring, and {\brg} in the ring.  However, generally the
equivalent widths of the emission lines in the ring are smaller than
e.g. in the rings of NGC 1097 (Paper I; Kotilainen et al. 2000) or NGC
1365 (see below).

\begin{figure}
  \includegraphics[height=8.5cm]{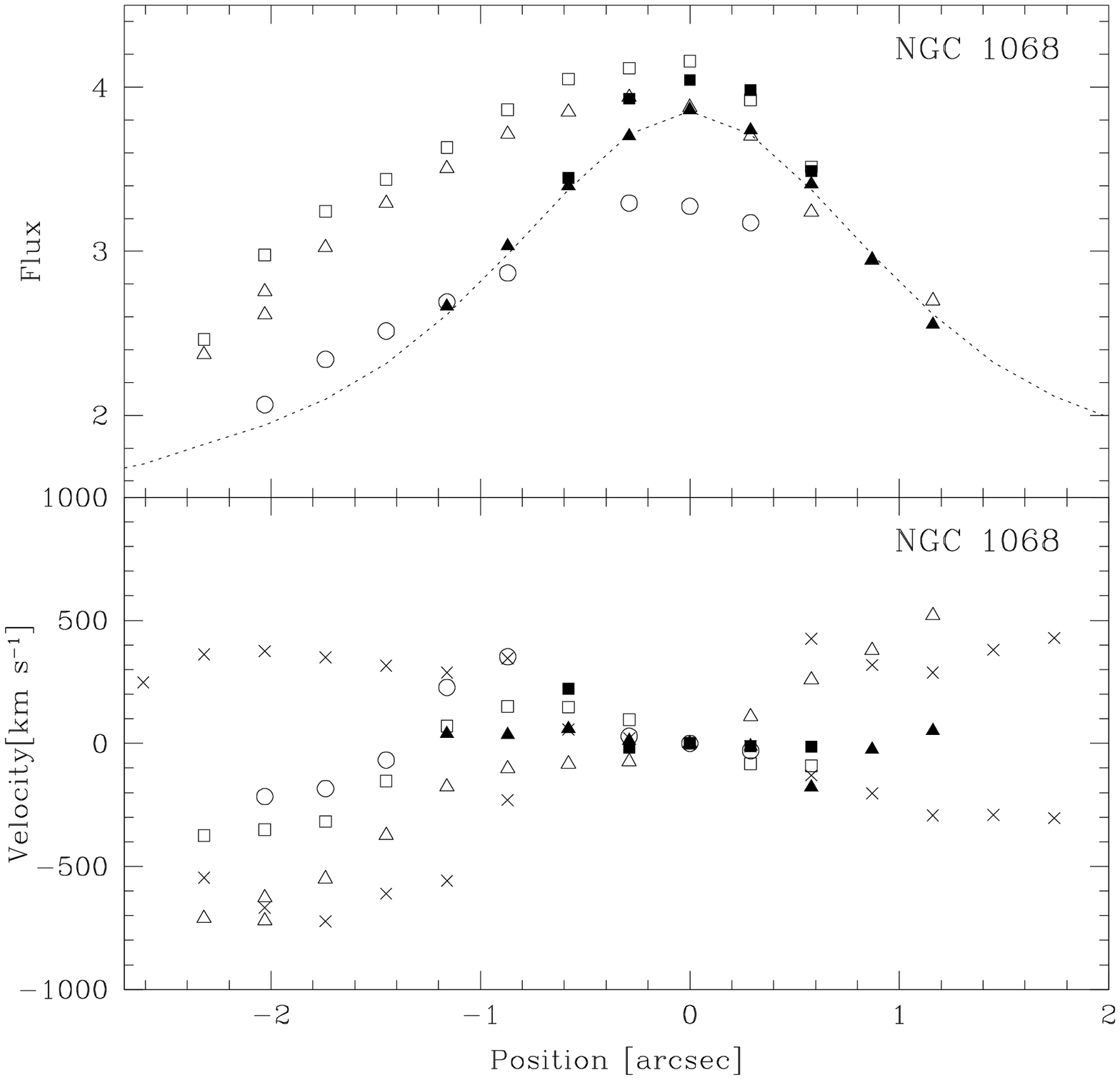}
  \caption{The spatial extent ({\em upper panel}) and velocity field 
    ({\em lower panel}) of the coronal {\sivi} 1.964 $\mu$m ({\em
      triangles}), {\caviii} 2.321 $\mu$m ({\em circles}) and {\sivii}
    2.483 $\mu$m line emission ({\em squares}) in NGC 1068.  Open
    symbols denote parallel to the cone and filled symbols perpendicular
    to it.  PSF is indicated with a dotted line. The two velocity
    components of {\iron} 1.644 $\mu$m are indicated with crosses.
    \label{n1068coro}}
\end{figure}

The kinematics of NGC 1068 has been discussed in detail by Schinnerer
et al. (2000) based on CO millimeter lines and by Alloin et al.
(2001), based on {\h2} emission.  The velocity curve of {\iron}
parallel to the cone is much steeper within the inner 3 arcsec (230
pc) than that of \h2. At larger radii, {\iron} seems to follow \h2
more closely. Furthermore, within the steep part of the velocity
curve, the kinematics of the {\iron} depends only on the position
along the slit. Thus it seems very likely that {\iron} is excited by
the outflowing material in the jet. In addition, parallel to the cone
{\iron} has two velocity components and thus the line profile changes
within the few inner arcsec (Fig.  \ref{n1068coro}).  Similar
structure is also visible in the {\brg} and coronal lines.

The nuclear \iron/{\brg} ratio = 4 (Table \ref{ratio}) is higher than
star formation models predict (e.g. Colina 1993). The 1--0 S
S(1)/{\brg} ratio = 0.4 is in agreement with the star formation origin
for H$_2$ as a result of UV heating from OB stars (Puxley et al.
1990). Whether the heating mechanism is thermal or fluorescence is not
clear, because no 2--1 S(1) or 1--0 S(0) emission was detected
in the nucleus.

The derived parameters, including the density and mass of the
molecular hydrogen both in the nuclear 1.4 arcsec aperture and
integrated within the starforming ring as traced by \brg, and the
spatial extent of different lines are given in Table 4.  In the
nuclear 1.4 arcsec aperture, the column density of the excited
molecular hydrogen is 1.3$\times$10$^{19}$ cm$^{-2}$ (corresponding to
$M_{\rmn{H_2}} = 1800$ M$_\odot$), assuming $T$ = 2000 K.  The FWHM
size of the nuclear {\h2} emission cannot be measured reliably as the
emission is weak with respect to the continuum.

Several IR coronal lines have previously been detected in NGC 1068
(e.g. Oliva \& Moorwood 1990).  All the detected coronal lines
({\sivi} 1.964 $\mu$m, {\caviii} 2.321 $\mu$m and {\sivii} 2.483
$\mu$m) are resolved parallel to the cone (Fig. \ref{n1068coro}), and
are more extended toward north ($\sim$2 arcsec) than south, in
agreement with the NICMOS image of {\sivi} emission by Thompson et al.
(2001).  The velocity curves of the coronal lines resemble the weaker
of the two {\iron} velocity components. Extended coronal line emission
can in principle only be excited by shocks in the interstellar medium
that are produced either by a jet or AGN-generated winds.

\subsection{NGC 1365}

NGC 1365 is a nearby ($z$ = 0.00546; distance = 21.7 Mpc) SB(s)b
galaxy in the Fornax cluster with a Seyfert 1.8 nucleus. NGC 1365 is
undergoing intense starburst activity in its bar concentrated
into a ring-like structure with a diameter of $\sim$15 arcsec (1.6
kpc) (e.g. Kristen et al. 1997; Stevens et al. 1999). The hollow
high-excitation outflow cone of NGC 1365 (PA = 130\degr; e.g. Kristen
et al. 1997) is parallel to the radio jet (Sandqvist et al. 1995).

\begin{figure}
  \includegraphics[width=8.5cm]{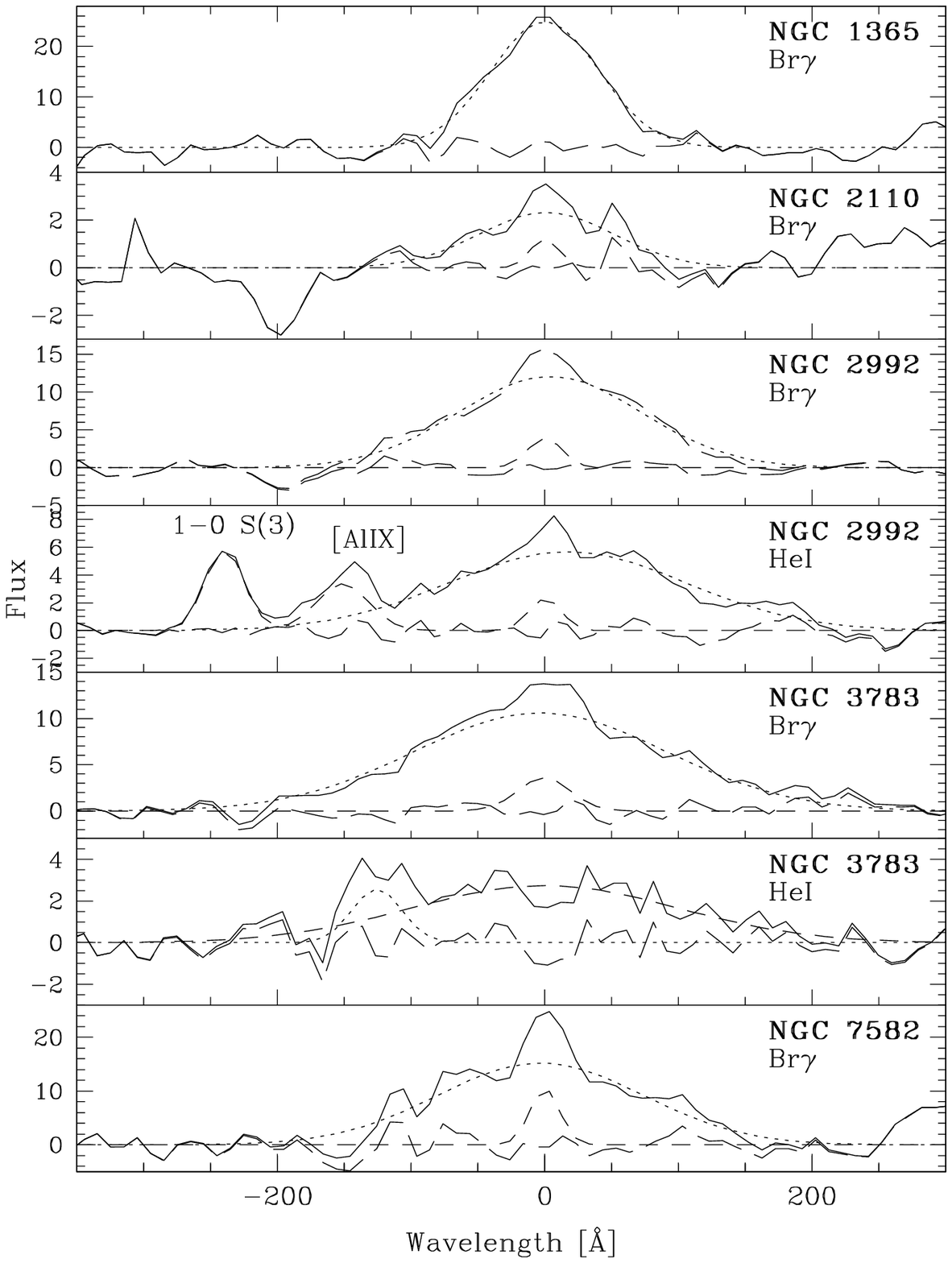}
  \caption{Decomposition of emission lines into narrow 
    ({\em short dashed line}) and broad ({\em dotted line}) components
    and the residual ({\em long dashed line}) of the fit for NGC 1365,
    NGC 2110, NGC 2992, NGC 3783 and NGC 7582. The continuum has been
    subtracted with a second order polynomial.
    \label{brg}}
\end{figure}

The $H$- and $K$-band nuclear spectra of NGC 1365 are shown in Fig.
\ref{allhk}, the extent of the main emission lines in Fig.
\ref{emission} and the spectra at different distances from the nucleus
in Fig. \ref{sp1}. The most intriguing feature in the spectra is that
unlike all the other galaxies in our sample, NGC 1365 does not exhibit
nuclear H$_2$ emission. This may be partly due to the strong
non-stellar continuum which dominates the $K$-band emission. In fact,
only {\he} 2.058 $\mu$m, several Brackett series lines, and weak
{\iron} 1.64 $\mu$m, are detected. {\brg} is broad with a FWHM of
$\sim$1190 \kms, and no evidence for the narrow component is seen
(Fig. \ref{brg}).

{\brg} 2.166 $\mu$m is the only extended nuclear emission line. It
decays smoothly towards the star forming regions ($\sim$6 arcsec; 600
pc), where other emission lines (e.g. \h2 and \he) become
visible. {\iron} is detected perpendicular to the ionization cone, but
not parallel to it, as there are no strong star forming regions
coincident with the slit in that direction. There is a fair
correlation between 1--0 S(1), \brg, {\iron} and {\he} in the star
forming ring, suggesting a common origin. This is further supported by
the fairly similar velocity curves in the ring (Fig. \ref{velocity}).

The velocity field of NGC 1365 is shown in Fig. \ref{velocity}. The
nuclear Br$\gamma$ emission is blueshifted by $\sim$220 {\kms} with
respect to the overall velocity curve. This blueshift is also observed
in the $H$-band Brackett lines, which are also broad ($\sim$1170
\kms).

No \h2 lines are detected in the nucleus. The 3$\sigma$ upper limit
for 1--0 S(1) emission in the nuclear 1.4 arcsec aperture is
8$\times$10$^{-16}$ \ergs giving an upper limit in an 1.4 arcsec
aperture of 60 M$_\odot$. The integrated mass of the excited hydrogen
is 290 M$_\odot$ perpendicular to the cone and $\sim$90 parallel to it.

Perpendicular to the ionization cone, there are three separate {\brg}
emission regions, the two innermost of which are associated with the
starburst ring. These regions are also visible in the $H$-band
Brackett lines, {\he} and {\iron}, which trace star formation, and
also in 1--0 S(1). The equivalent width of {\brg} reaches up to 70
{\AA}, much larger than e.g. in the ring of NGC 1097
(Kotilainen et al. 2000; $\sim$15 \AA), implying recent star
formation. Parallel to the cone there are no separate {\brg} knots,
but the ring shows off as a plateau in the spatial line emission
profile. The extinction derived from the Br 10--4/Br$\gamma$ and Br
11-4/Br$\gamma$-ratios ($A_{\rmn{K}}$ = 0.67 and 0.69, respectively)
is identical to that based on the continuum colour
($A_{\rmn{K}}$ = 0.68).  Notwithstanding the existence of a strong UV
radiation field as evidenced by the Brackett lines, {\h2} appears to
be thermally excited, as the 2--1 S(1)/1--0 S(1) ratio is 0.16\PM0.05
($T_{\rmn{vib}} = 2600 \pm 400$ K).

\subsection{NGC 2110\label{section2110}}

NGC 2110 is a nearby ($z$ = 0.00779; distance = 31.0 Mpc) SAB0$^{-1}$
galaxy with a Sy2 nucleus. HST {\oiii} imaging shows a narrow 1 arcsec
long jet-like feature (cone) at PA = 340\degr, and a weaker feature at
PA = 160\degr (Mulchaey et al. 1994).  The hard X-ray emission of NGC
2110 (Malaguti et al. 1999) is highly absorbed and indicates an
obscured Sy1 nucleus.  Storchi-Bergmann et al. (1999) found extended
{\h2} 1--0 S(1) emission, probably excited by central X-rays.

The nuclear $H$- and $K$-band spectrum of NGC 2110 (Fig. \ref{sp1}) is
different from the more powerful Sy2s such as NGC 1068 (this paper)
and NGC 5128 (Paper I), appearing to have more contribution from the
stellar continuum. {\iron} 1.644 $\mu$m is by far the strongest line
in the nucleus, where other detected lines are various {\h2} lines and
{\brg} 2.166 $\mu$m. {\brg} is rather weak, but has a faint broad
component (FWHM $\sim$ 2000 \kms; Fig. \ref{brg}).  No
coronal lines are detected.

NGC 2110 has been previously observed spectroscopically in the NIR by
e.g. Storchi-Bergmann et al. (1999).  The presence of a broad
component to the recombination lines has been under debate as
Veilleux et al. (1997) found some evidence in Pa$\beta$
1.282 $\mu$m, but Storchi-Bergmann et al. (1999) did not detect this.
We found the FWHM of {\brg} to be $\sim$1550 \kms, in fair agreement
with Veilleux et al. (1200 \kms).

1--0 S(1) is the most extended line, followed by \iron, but even 1--0
S(1) can only be traced up to 4--5 arcsec ($\sim$700 pc) distance from
the nucleus parallel to the cone and 1.5 arcsec ($\sim$300 pc)
perpendicular to it with no separate emission regions. Surprisingly,
Br$\gamma$ is not extended, which excludes star formation as the
origin of the extended {\iron} emission.  The total (nuclear) extent
of {\iron} emission is $\sim$4 arcsec (0.6 kpc) parallel to the cone,
and $\sim$2 arcsec (300 pc) perpendicular to it. {\iron} is spatially
resolved in the nucleus and is more extended parallel to the cone.

The velocity curve of {\iron} parallel to the cone is flatter than that of 
{\h2} close to the nucleus, in agreement with previous determinations (e.g.
Knop et al. 2001). Further out the velocity curves of {\h2} and 
{\iron} are in better agreement. Perpendicular to the cone the curves are
flatter than parallel to the cone.

The dereddened {\iron}/{\brg} ratio is $\sim$8, one of the highest in
our sample, and {\iron}/{\brg}$_{narrow}$ = 80. Thus, considering the
different morphologies of {\iron} and {\brg}, {\iron} is likely to be
X-ray excited. Furthermore, there is a fair correlation between {\em
Chandra} soft X-ray images (Weaver 2001) and the {\iron} emission: the
X-ray emission in NGC 2110 is more extended parallel to the cone than
perpendicular to it, and more extended towards north than south. X-ray
excitation (Maloney, Hollenbach \& Tielens 1996) can also explain the
{\h2} emission in the nucleus, assuming that almost all of the gas is
dense ($n=10^5$ cm$^{-3}$) and that there is a high ratio of the local
X-ray energy deposition rate per particle to the gas density. Outside
of the nucleus, {\h2} may also be X-ray excited as 1--0 S(1)/{\iron}
ratio produced in X-ray excited gas depends strongly on the density of
the gas and the intensity of the radiation, but this does not seem
likely as the velocity fields of {\iron} and {\h2} are quite
different.

Assuming $T_{\rmn{vib}} = 2000$ K, we derive $N_{\rmn{H_2}} =
4.3\times10^{17}$ cm$^{-2}$ (corresponding to $M_{\rmn{H_2}} \simeq
250$ M$_\odot$) within the nuclear aperture. The total integrated
$M_{\rmn{H_2}}$ is larger parallel to the cone than perpendicular to
it (220 vs. 190 M$_\odot$) and the FWHM of nuclear {\h2} emission is
larger parallel to the cone than perpendicular to it (320 pc vs. 230
pc).

\begin{table*}
  \centering
  \begin{minipage}{180mm}
    \caption{The mass and the column density of the excited {\h2} in the 
      nucleus (1.4 arcsec aperture), the nuclear spatial size corrected for the seeing (FWHM;
      Section 2) and
      the integrated {\h2} mass within the starforming regions and the
      nuclear spatial extent of the lines, both parallel to the cone
      ($\parallel$) and perpendicular to it ($\perp$). The nuclear
      extent given in parsecs is in the plane of the galaxy and other
      values are in the plane of the sky. The integrated masses and
      the spatial extents are given over the whole galaxy for NGC 4945
      and NGC7582. Star $\star$ indicates that the central peak is
      significantly off-nuclear. \label{prm}}
    \begin{tabular}{lllllllllllllll}
      \hline
                                          &1068   &1097   &1365  &1386   &1566   &2110  &E428   &2992 &3081   &3227  &3783   &4945  &5128  &7582 \\
      Nucleus                             &Sy2    & Sy1   &Sy1.8 &Sy2   &  Sy1   &Sy2   & Sy2   & Sy2 & Sy2   &Sy1.5 & Sy1   &Sy2   &Sy2   & Sy2 \\\\
      Paper                               & II    & I     & II   & I     & I     & II   & II    & II  & II    & I    & II    & I    & I    & II  \\
      Broad \brg                          &no     &...    &yes   &yes    &yes    &yes   &no     &yes  &no     &yes   &yes    &no    &no    &yes  \\
      M(\h2) [M$_\odot$]                  &1800   &25     &$<$60 &17     &42     &250   & 100   & 500 & 110   &280   & 290   &55    &32    & 220 \\
      N(\h2) [10$^{17}$ cm$^{-2}$]        &130    &1.5    &$<$2.1&3.7    &1.9    &4.3   & 3.5   & 8.8 & 1.8   &22    & 3.2   &80    &44    & 7.8 \\
      M(\h2) $\parallel$ [M$_\odot$]      &2200   &90     &90    &50     &90     &220   & 200   & 350 & 80    &310   & 290   &70    &30    & 320 \\
      M(\h2) $\perp$ [M$_\odot$]          &2100   &60     &290   &30     &110    &190   & 160   & 560 & 160   &350   & 270   &150   &30    & 480 \\
      FWHM {\h2} $\parallel$ [pc]         &...    &99     &...   &$<$20  &65     &320&140$\star$& 120 & 210   &58    & ...   &28    &42    & ... \\
      FWHM {\h2} $\perp$ [pc]             &...    &108    &...   &39     &82     &230   & 150   & 250 & 160   &84    & ...&57$\star$&35    & ... \\
      FWHM {\iron} $\parallel$ [pc]    &100$\star$&...    &...   &$<$20  &$<$30  &220&160$\star$& 80  & ...   &50    & 130   &50    &$<$10 & 310 \\
      FWHM {\iron} $\perp$  [pc]          &80     &...    &...   &$<$20  &$<$30  &100   &130    & 140 & ...   &40    &$<$100 &50    &$<$10 & 160 \\
      Extent {\h2} $\parallel$ [\arcsec]  &5.4    &3.0    &...   &4.2    &2.0    &4.5   & 4.9   & 1.0 & 1.5   &6.0   & 0.6   &17    &10    & 2.3 \\
      Extent {\h2} $\perp$ [\arcsec]      &4.1    &0.9    &...   &1.0    &2.9    &1.5   & 1.5   & 6.7 & 1.0   &7.5   & 0.6   &36    &16    & 6.7 \\
      Extent {\iron} $\parallel$ [\arcsec]&3.2    &...    &...   &0.8    &0.5    &4.1   & 2.8   & 1.0 & 0.6   &1.6   & 0.6   &4.8   &1.2   & 1.9 \\
      Extent {\iron} $\perp$ [\arcsec]    &5.4    &...    &...   &0.8    &0.5    &1.9   & 1.9   & 6.2 & 0.6   &2.0   & 0.6   &7.2   &1.2   & 3.6 \\
      Extent {\brg} $\parallel$ [\arcsec] &5.4    &...    &3.0   &0.8    &0.5    &...   & 2.3   & 1.9 & 1.9   &2.3   & 1.0   &3.1   &1.1   & 1.9 \\
      Extent {\brg} $\perp$ [\arcsec]     &2.3    &...    &4.0   &0.8    &1.1    &...   & 1.0   & 2.3 & 0.6   &1.3   & 0.6   &7.8   &0.6   & 3.2 \\
      Extent {\h2} $\parallel$ [pc]       &450    &340    &...   &410    &210    &730   & 530   & 470 & 240   &600   & 130   &1600  &240   & 580 \\
      Extent {\h2} $\perp$ [pc]           &320    &80     &...   &190    &280    &300   & 270   &1300 & 200   &860   & 130   &1200  &720   & 720 \\
      Extent {\iron} $\parallel$ [pc]     &270    &...    &...   &80     &50     &660   & 300   & 480 & 90    &160   & 130   &420   &30    & 480 \\
      Extent {\iron} $\perp$ [pc]         &420    &...    &...   &150    &50     &380   & 340   &1200 & 120   &230   & 130   &230   &50    & 390 \\
      Extent {\brg} $\parallel$ [pc]      &450    &...    &590   &80     &50     &...   & 250   & 900 & 250   &230   & 220   &270   &30    & 480 \\
      Extent {\brg} $\perp$ [pc]          &180    &...    &430   &150    &100    &...   & 180   & 450 & 120   &150   & 130   &250   &30    & 340 \\
      T$_{vib}$ [10$^3$ K]                &$<$5.3 &$<$2.6 &...   &$<$2.5 &$<$2.7 &3.3   &$<$2.6 & 2.6 &$<$2.3 &2.0   &$<$4.8 &2.2   &$<$1.8& 2.9 \\
      \hline
    \end{tabular}
  \end{minipage}
\end{table*}

\subsection{ESO 428-G14}

ESO 428-G14 is a nearby ($z$ = 0.00544; distance = 21.6 Mpc)
SA:(l)0$^+$ galaxy with a Sy2 nucleus (Bergvall, Johansson \& Olofsson
1986).  It has a bent radio jet (Ulvestad \& Wilson 1989), and an
{\oiii} cone along the jet (Wilson \& Baldwin 1989). The NLR of ESO
428-G14 consists of many thin strands closely related to the two-sided
jet (Falcke et al. 1996).

\begin{figure*}
  \includegraphics[]{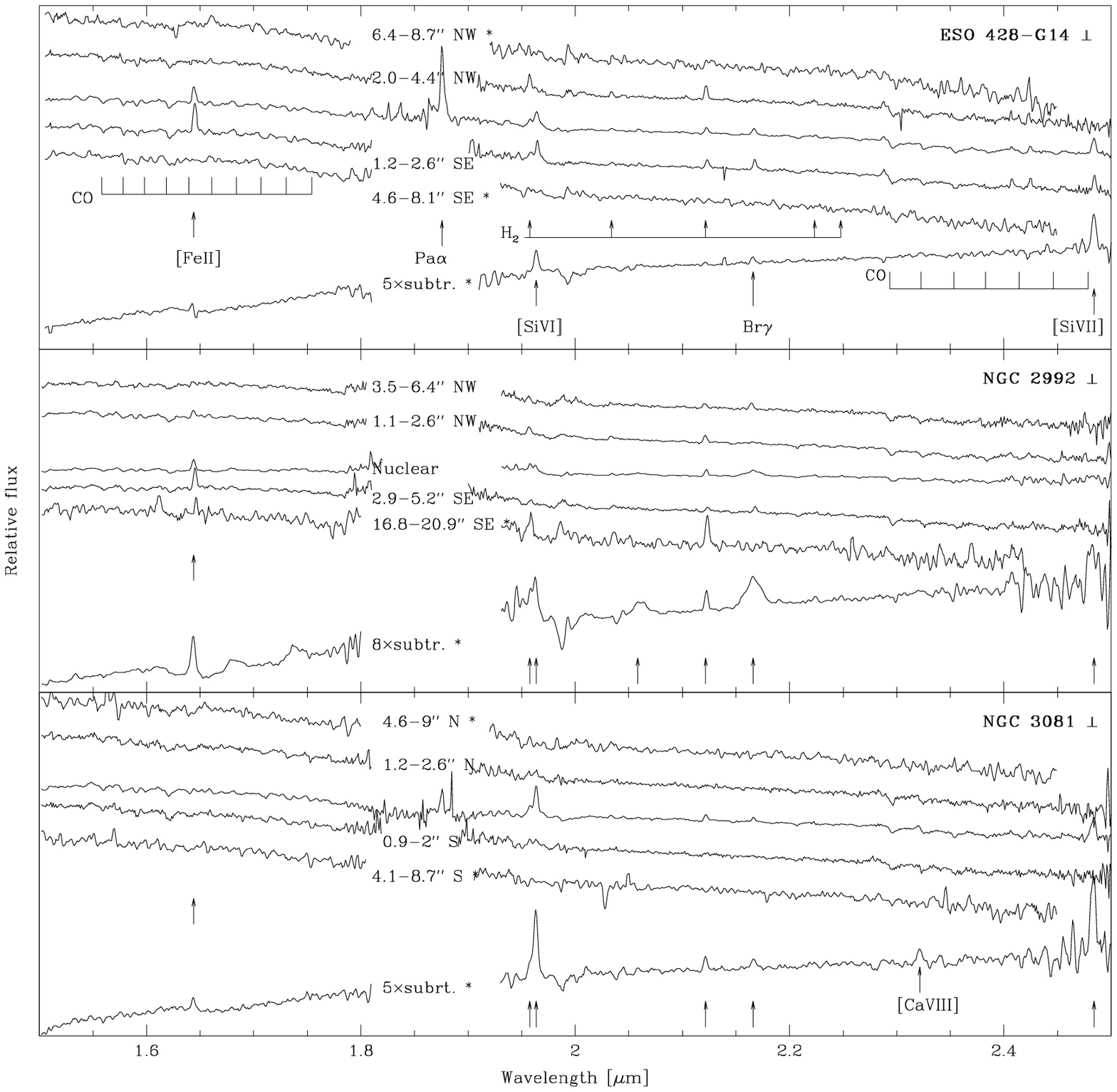}
  \caption{The 1.5--2.5 $\mu$m spectra of ESO 428-G14 ({\em top}), NGC 2992 
    and NGC 3081 ({\em bottom}) perpendicular to the cone at different
    positions along the slit. For symbols, see Fig. \ref{sp1}.
    \label{sp2}}
\end{figure*}

The $H$- and $K$-band spectra of ESO 428-G14 are shown in Fig.
\ref{allhk} and Fig. \ref{sp2}.  The strongest nuclear line is {\iron}
1.644 $\mu$m, followed by {\brg} 2.166 $\mu$m, various {\h2} lines,
and {\he} 2.058 $\mu$m.  Two coronal lines {\sivi} 1.964 $\mu$m and
{\sivii} 2.483 $\mu$m were also detected. 

The nuclear {\h2}, {\iron} and {\brg} emission lines are all spatially
resolved. 1--0 S(1) is the most extended line,
extending up to $\sim$7 arcsec (0.7 kpc) parallel to the cone, but
only up to $\sim$2 arcsec (210 pc) perpendicular to it. Also, both
Br$\gamma$ and {\iron} are more extended parallel to the cone (up to $\sim$3
arcsec; 320 pc) than perpendicular to it ($\sim$2 arcsec; $\sim200$ pc). The
morphologies of H$_2$, {\brg} and {\iron} are all similar parallel to
the cone: the central peak is displaced 0.5 arcsec to NE and a separate
peak is detected $\sim$1.3 arcsec SW of the continuum nucleus.
Br$\gamma$ is narrow with no detectable broad component.

The velocity field of ESO 428-G14 is shown in Fig. \ref{velocity}.
{\iron} is blueshifted by $\sim$70 {\kms} across the nuclear region.
\h2 and {\brg} appear to have similar dynamics, but in the NE peak {\brg} is also
blueshifted with respect to H$_2$. Thus it seems that the
extended {\h2} and {\brg} smission share a same origin, probably in the star
forming clouds.

The nuclear \iron/{\brg} ratio = 2.6 is higher than models suggest for
star forming regions. The ratio is nearly constant parallel to the
cone up to $\sim$2 arcsec from the nucleus. Perpendicular to the cone,
the ratio is lower ($\sim$1.2) toward northwest. The 1--0 S(1)/{\brg}
ratio is also fairly constant, 0.6--0.9, across the nucleus. 

Based on the velocity field and line ratios, {\h2} is most likely
thermally excited by UV radiation from OB stars. Since {\iron} is
blueshifted in the nucleus and is more extended parallel to the cone
than perpendicular to it, {\iron} is likely to be shock-excited by the
winds in the cone.  In the NE peak shocks also seem to contribute to
the {\brg} emission.

No 2--1 S(1) is detected, and the upper limit to the 2--1 S(1)/1--0
S(1) ratio is in agreement with thermal excitation of H$_2$. Assuming
$T_{\rmn{vib}} = 2000$ K, we derive $N_{\rmn{H_2}} = 3.5\times10^{17}$
cm$^{-2}$ (corresponding to $M_{\rmn{H_2}} \simeq 100$ M$_\odot$)
within the nuclear aperture.  The integrated mass of the excited
hydrogen is 200 M$_\odot$ parallel to the cone and 160 M$_\odot$
perpendicular to it. The FWHM size of the nuclear {\h2} emission is
similar in both directions ($\sim$150 pc).

Of the detected coronal lines, {\sivi} 1.964 $\mu$m is spatially
resolved parallel to the cone (Fig. \ref{emission}), and is more
enhanced toward southwest than northeast. {\sivii} 2.483 $\mu$m is
also spatially resolved and similarly extended, though its
determination suffers from the much higher thermal background noise.
The {\sivi} emission within the innermost 2 arcsec is composed of a
resolved nuclear source with FWHM$\sim$ 1.6 arcsec (170 pc) and an
additional, decaying extended component towards southwest.
Furthermore, after subtracting the underlying CO lines, a marginal
{\caviii} 2.321 $\mu$m line appears at a 2.8$\sigma$ level with flux
0.42\PM0.15 10$^{-15}$ \ergs.

\subsection{NGC 2992}

NGC 2992 is a nearby ($z$ = 0.00771; distance = 30.6 Mpc) almost
edge-on peculiar Sa galaxy with a Sy2 nucleus, interacting with NGC
2993 at 2.9 arcmin (26 kpc) distance toward SE. A prominent dust lane
roughly N-S along the major axis of the galaxy splits the nuclear
region into two components.  NGC 2992 has an unusually extensive cone
emerging almost perpendicularly from the plane of the galaxy up to
$\sim$4 kpc ($\sim$25 arcsec) distance in SE-NW direction (Allen et al. 1999), with the
SE cone brighter than the NW cone.  The radio structure of NGC 2992
has major axis PA = 160\degr, with a one-sided extension along PA =
130\degr, and at smaller scales a ``figure of 8'' shape at PA =
160\degr (e.g.  Ulvestad \& Wilson 1984).

The $H$- and $K$-band spectra of NGC 2992 are shown in Fig.
\ref{allhk} and Fig. \ref{sp2}. The strongest emission line in the
nucleus is the broad {\brg} 2.166 $\mu$m.  Other nuclear lines are
{\iron} 1.644 $\mu$m, {\he} 2.058 $\mu$m, coronal lines and various
{\h2} lines, including 2--1 S(1) 2.248 $\mu$m.  NGC 2992 has been
previously observed spectroscopically in the $J$$H$$K$-bands in
January 1999 also with SOFI and 1 arcsec slit by Gilli et al (2000).
Compared to them, the emission lines have become brighter and {\brg}
much broader (2530 \kms) than their Pa$\alpha$.  Furthermore a
relatively strong {\alix} line at 2.043 $\mu$m has appeared.

The most extended line is 1--0 S(1), which can be traced up to 7
arcsec (1.1 kpc) distance perpendicular to the cone but only up to 1
arcsec (220 pc) parallel to it. Both Br$\gamma$ and {\iron} are more
extended ($\sim$4 arcsec; 600 pc) parallel to the cone than 1--0 S(1).
In addition, toward southwest {\h2} emission is detected associated
with the bridge connecting NGC 2992 to NGC 2993. This emission is
likely to be shock-excited, as 2--1 S(1) emission is not detected.
There are also separate {\brg} regions perpendicular to the cone 5
arcsec (760 pc) from the nucleus, coinciding with the bar or spiral
arms (Fig. \ref{broad}).

The velocity field of NGC 2992 is shown in Fig. \ref{velocity}. The
velocity curve of {\iron} is blueshifted by 70 \kms. However, the
velocity curve of {\brg} seems to follow more closely \h2 than \iron.

The 1-0 S(1)/{\brg} ratio is quite low, 0.28, suggesting thermal UV
heating.  The {\iron}/{\brg} ratio is 0.93 and
{\iron}/{\brg}$_{narrow}$ = 16.  Considering that {\iron} is
blueshifted in the nucleus, it is plausible that {\iron} is excited by
shocks.  The narrow {\brg} emission may also originate in shocks, as
the narrow {\brg}/1--0 S(1) ratio is $\sim$0.2, but as {\brg} is
kinematically more closely coupled with {\h2} than {\iron}, the
shocked regions producing {\brg} are not likely to coincide with
{\iron} regions.  In addition to Br$\gamma$, also {\he} is broad (2200
\kms) (Fig. \ref{brg}).

The dereddened 2--1 S(1)/1--0 S(1) ratio 0.18 (corresponding to
$T_{\rmn{vib}} = 2600$ K) is in agreement with thermal excitation of
{\h2}.  The density of the excited molecular hydrogen is
$N_{\rmn{H_2}} = 8.8\times10^{17}$ cm$^{-2}$, corresponding to
integrated gas mass of $\sim$500 M$_\odot$. 
The FWHM size of the nuclear H$_2$ emission is smaller
parallel to cone (120 pc) than perpendicular to it (250 pc).  

Three coronal lines ({\sivi} 1.964 $\mu$m, {\sivii} 2.483 $\mu$m and
{\alix} 2.043 $\mu$m) are detected in NGC 2992.  In addition, there
appears to be a fairly broad (80 \AA) {\caviii} 2.321 $\mu$m line
detected at 2.5$\sigma$ level.  The coronal lines are weaker than
those in NGC 1068 and NGC 3081 (this paper), and it is not possible to
discuss whether the lines are spatially resolved.

\subsection{NGC 3081}

NGC 3081 is a nearby ($z$ = 0.00795; distance = 31.6 Mpc) 
Sy2 galaxy with four rings (e.g. Buta \& Purcell 1998), a large-scale bar 
at PA = 66\degr, and a nuclear bar at PA = 120\degr 
(Storchi-Bergmann et al. 1996). H$\alpha$ imaging shows a ring of 
{\hii} regions with major axis perpendicular to the large-scale bar 
(Storchi-Bergmann et al.).

The $H$- and $K$-band spectra of NGC 3081 are shown in Fig.
\ref{allhk} and Fig. \ref{sp2}. The coronal lines {\sivi} 1.964 $\mu$m
and {\sivii} 2.484 $\mu$m are by far the strongest emission lines
detected in the nuclear region.  Also the coronal {\caviii} 2.321
$\mu$m and {\alix} 2.043 $\mu$m lines, various {\h2} 1--0
S- and Q-transition lines, {\he} 2.058 $\mu$m, narrow \brg, and a weak
{\iron} 1.644 $\mu$m are detected.

The most extended line is {\sivi} followed by {\brg} and 1--0 S(1)
2.122 $\mu$m, which both extend $\sim$ 2 arcsec (310 pc; Fig.
\ref{emission}). \brg, 1--0 S(1) and {\iron} are very narrow, comparable to the
instrumental width of the lines ($\sim$30--32 \AA). Also {\iron} is
not spatially resolved.

The {\iron}/{\brg} ratio is 0.64, one of the lowest in the sample, and
can be produced by star formation.  The 1--0 S(1)/{\brg} ratio is 1.6,
indicating shocks as excitation mechanism also for {\h2}.  The upper
limit to the nuclear 2--1 S(1)/1--0 S(1) ratio is 0.13, while the 1--0
S(0)/1--0 S(1) and 1--0 S(2)/1--0 S(1) ratios are 0.13 and 0.57,
respectively, clearly ruling out UV fluorescence, and suggesting
shocks to be responsible for the excitation of {\h2}.  In the nuclear
1.4 arcsec aperture, the gas density $N_{\rmn{H_2}} =
1.8\times10^{17}$ cm$^{-2}$, corresponding to gas mass of only 110
M$_\odot$. The FWHM size of the nuclear {\h2} emission is larger
parallel to the cone (210 pc) than perpendicular to it (160 pc). The
integrated mass of excited {\h2} is 80 M$_\odot$ parallel to the cone
and 70 M$_\odot$ perpendicular to it.

In contrast with the other galaxies in the sample with 
circumnuclear rings (e.g. NGC 1097, NGC 1365 and NGC 4945), the rings
in NGC 3081 are not prominent in the spectra. The effective slit
length covers only the nuclear ring (PA$\sim$120\degr; major axis
$\sim$12 arcsec; 1.9 kpc) and partially the inner ring ($\sim$75\degr;
70 arcsec; 11 kpc). In the nuclear ring, Br$\gamma$ 2.166 $\mu$m is
only visible after smoothing the spectra, and no emission lines from
the inner ring can be detected.

The coronal {\sivi} line is more extended parallel to the cone (FWHM
2.2 arcsec; 350 pc) than perpendicular to it (1.2 arcsec; 200 pc).
Perpendicular to the cone the emission peak is slightly off-nuclear.
This extended emission cannot be detected in the {\sivii} line due to
the much higher background noise this line suffers.

\subsection{NGC 3783}

NGC 3783 is an (R')SB(r)a galaxy with a Sy1 nucleus. It is the most distant 
($z$ = 0.00973; distance = 38.6 Mpc) galaxy in our sample, but one of the 
closest Sy1s, and thus has been extensively studied (e.g. Maran et al. 1996).
It has an inner ring (diameter $\sim$35 arcsec; 6.7 kpc) with star
formation activity circling a strong stellar bar (PA = 163\degr), and
thin tightly wound spiral arms emerging from this ring (e.g.
Mulchaey et al. 1997). NGC 3783 has extremely broad Balmer and He lines which 
are blueshifted relative to the systemic velocity (Evans 1988). 

\begin{figure*}
  \includegraphics[]{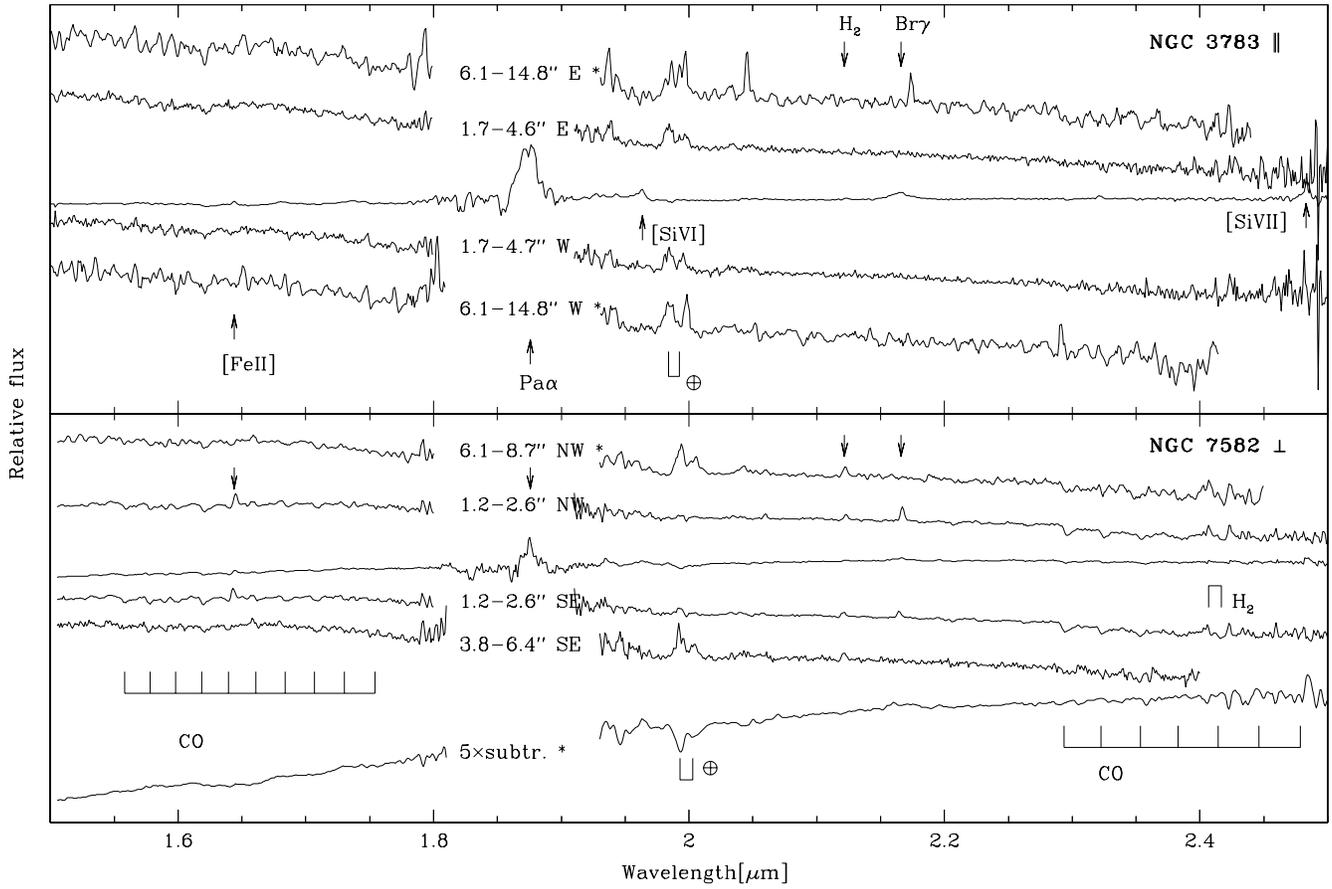}
  \caption{The 1.5--2.5 $\mu$m spectra of NGC 3783 ({\em top}) and NGC 7582 
    ({\em bottom}) at different positions along the slit. For symbols,
    see Fig. \ref{sp1}.
    \label{sp3}}
\end{figure*}

\begin{figure*}
  \begin{minipage}[t]{0.4\linewidth}
    \includegraphics[height=7cm]{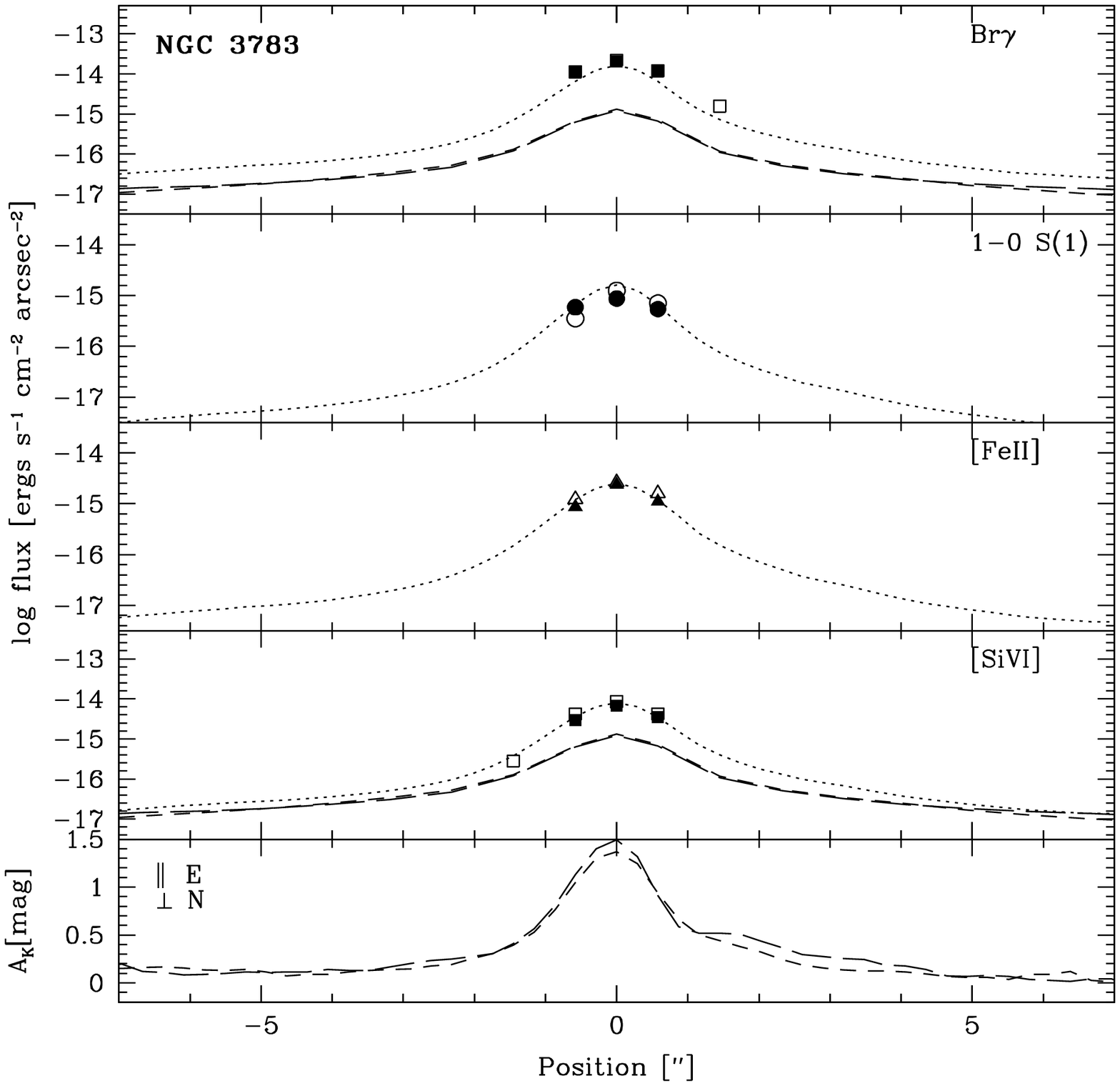}
  \end{minipage}
  \begin{minipage}[t]{0.4\linewidth}
    \includegraphics[height=7cm]{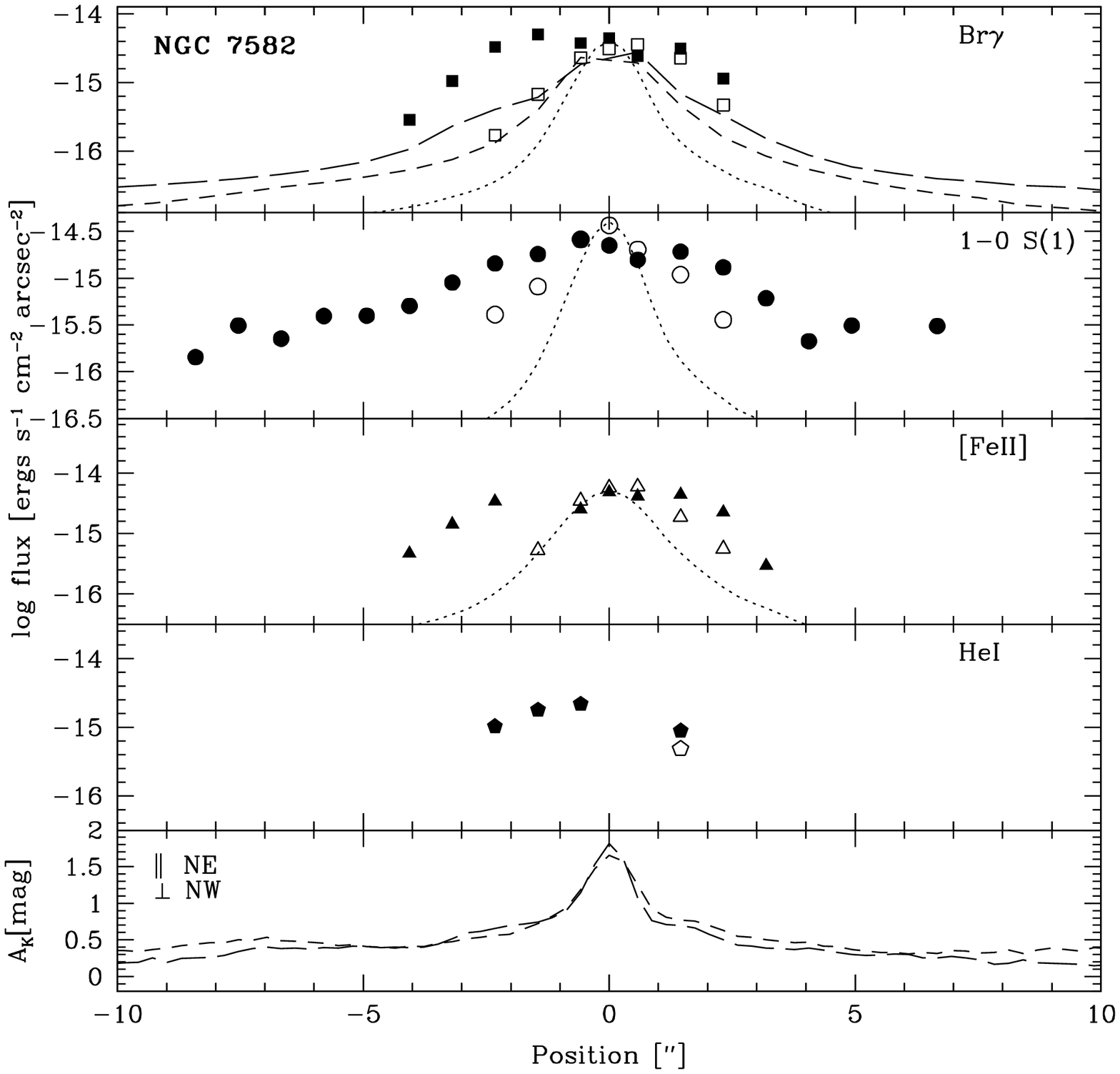}
  \end{minipage}
  \caption{The observed line emission in NGC 3783 ({\em left}) and NGC 7582 
    ({\em right}). Symbols are as in Fig. \ref{emission}. 
    \label{emission2}}
\end{figure*}

Non-stellar continuum appears to dominate the overall spectrum of NGC
3783 (Fig. \ref{sp3}). The strongest line within the nuclear aperture
is the broad ($\sim$ 2630 \kms) {\brg}. Other lines detected are
1--0 S(1), {\iron}, {\he}, $H$-band Brackett lines and three coronal lines.
Similarly to the case of NGC 3081, NGC 3783 is rather distant and
faint and each line can only be traced up to $\sim$1 arcsec (200 pc)
from the nucleus (Fig. \ref{emission2}).  Thus, no kinematical
analysis can be presented for NGC 3783.

Similarly to the case of NGC 2992, in addition to broad \brg, a faint
broad {\he} line is detected (Fig. \ref{allhk}). This line also
appears to be double-peaked in contrast to {\brg}(Fig. \ref{brg}). The width of {\he}
($\sim$3700 \kms) is broader than that of \brg, and the peaks are
separated by $\sim$1000 \kms.  In the $H$-band, other Brackett transition lines are
detected (Br 10-4 1.737 $\mu$m, Br 11-4 1.681 $\mu$m and Br 12-4 1.641
$\mu$m).  Their ratios against {\brg} indicate that the BLR is
unobscured.

The upper limit to the 2--1 S(1)/1--0 S(1) ratio is $\sim$0.3
($T_{\rmn{vib}} <$ 4800 K), while for fluorescent emission the
expected ratio is $\sim$0.5. Thus we cannot exclude the contribution
from fluorescent \h2 emission. The density of the excited {\h2} in the
nucleus $N_{\rmn{H_2}} \simeq 3.2\times10^{17}$ cm$^{-2}$
(corresponding to $M_{\rmn{H_2}}$ $\simeq$ 290 M$_\odot$).

Three coronal lines are detected in NGC 3783: {\sivi} 1.964 $\mu$m,
{\caviii} 2.321 $\mu$m and {\sivii} 2.483 $\mu$m. {\sivi} and {\sivii}
are both spatially resolved in the nucleus but no additional extended
emission is detected. The FWHM size of the nuclear emission source
(corrected for the instrumental resolution) is $\sim$0.7 arcsec with
no significant difference between the directions. In addition, there
appears to be a faint {\alix} line detected at a 2.5$\sigma$ level
blended with the broad {\he} line, but higher S/N spectra are required
to confirm its presence.

\subsection{NGC 7582}

NGC 7582 is a nearby ($z$ = 0.00540; distance = 20.9 Mpc)
(R'$_1$)SB(s)ab galaxy in the Grus quartet with a Sy2 nucleus. Optical
broad-band images (e.g. Malkan, Gorjian \& Tam 1998) show a disturbed
morphology with an optically obscured nucleus, a red star forming ring
(diameter 2 arcsec) and a dust lane crossing the nuclear region. In
the NIR, the galaxy appears smoother, and in the $L$-band the nucleus
dominates albeit the ring remains visible (Prieto, Reunanen \&
Kotilainen 2002).  Optical spectropolarimetry does not reveal a hidden
BLR (Heisler et al. 1997), and even hard X-ray emission is attenuated
by large column density (Warwick et al. 1993), indicating that an
edge-on torus blocks even the scattered light from the nucleus.
However, the nuclear emission is variable, and the broad hydrogen
recombination line has been detected by Sosa-Brito et al. (2001). The
sharp-edged {\oiii} outflow cone of NGC 7582 (PA = 250\degr) was
detected by Storchi-Bergmann \& Bonatto (1991).

The $H$- and $K$-band spectra are shown in Fig. \ref{sp3}.  The
strongest nuclear line is \brg, while the other lines detected are
1--0 S(1) and other \h2 lines, {\iron} and an unidentified emission
feature at 2.195 $\mu$m, which remains visible after subtracting the
continuum. Br$\gamma$ can be decomposed into broad ($\sim$ 3000 \kms)
and narrow (420 \kms) components (Fig. \ref{brg}). The broad component
has an additional blue wing and is therefore not perfectly fitted by a
simple two-component fit. Two coronal lines ({\sivi} and {\sivii}) are
also detected.

1--0 S(1) is the most extended line (Fig. \ref{emission2}) and is more
extended perpendicular to the cone (8 arcsec; 900 pc) than parallel to
it (2.5 arcsec; 270 pc). 1--0 S(1) emission declines smoothly, but the
effect of the ring on the {\h2} emission is also notable, especially
perpendicular to the cone.  The ring is more prominent in {\brg} and
{\iron}, which are detected parallel to the cone up to 4 arcsec (430
pc) in northeast and 2.5 arcsec in southwest and $\sim$2 arcsec
perpendicular to the cone.  The emission lines are narrow, comparable
to or slightly larger than the instrumental resolution. {\he} is also
strong in the ring, in agreement with relatively recent star
formation.

The velocity field of NGC 7582 is shown in Fig. \ref{velocity}.
Parallel to the cone the velocity curves are relatively flat, and
similar for the various emission lines. {\iron} is
systematically blueshifted with respect to \h2.  The ratio of
\iron/narrow {\brg} = 3.1 is larger than models predict for
star forming regions, but in agreement with shocks or X-rays.

The ratio of 2--1 S(1)/1--0 S(1) = 0.20\PM0.06 in the nucleus in
agreement with thermal excitation of \h2 ($T_{\rmn{vib}} =$ 2900\PM500
K). Assuming $T_{\rmn{vib}}=2000 $ K, the density of {\h2} is
7.8$\times10^{17}$ cm$^{-2}$, corresponding to a gas mass of 220
M$_\odot$. The integrated mass of the excited hydrogen is 320
M$_\odot$ parallel to the cone and 480 M$_\odot$ perpendicular to it
across the galaxy. The nuclear H$_2$ emission is too weak to
accurately determine the FWHM size of the nucleus.

The coronal lines {\sivi} 1.964 $\mu$m and {\sivii} 2.483 $\mu$m are
 are weak, and it is not possible to
reliably measure the size of the coronal emission region. However, the
detection of coronal lines supports the presence of AGN in NGC 7582,
as does the detection of broad {\brg}.

\section{Discussion}

We have presented NIR 1.5--2.5 $\mu$m long-slit spectra of eight
Seyfert galaxies with an ionization cone. Combined with similar data
on six Seyferts from Paper I, the total sample size is 14 Seyferts
(three Sy1s, two intermediate Seyferts and nine Sy2s). The discussion
below is based on this total sample. The strongest emission line is
{\brg} 2.166 $\mu$m in 43 \% of the galaxies (two Sy1s, two Sy2s and
two intermediate Seyferts), {\iron} 1.644 $\mu$m in 21 \% (three Sy2s), 1-0
S(1) 2.122 $\mu$m in 14\% (one Sy1, one Sy2) and a coronal line
(either {\sivi} 1.964 $\mu$m or {\sivii} 2.484 $\mu$m) in 21 \% (three
Sy2s).  If a broad {\brg} line is detected, it is the strongest line
in 5/8 (62 \%) galaxies, while {\iron} remains the strongest line in
NGC 2110 and NGC 3227 and {\sivii} in NGC 1386.

\subsection{Br$\gamma$ emission}

\begin{figure}
  \includegraphics[width=8.5cm]{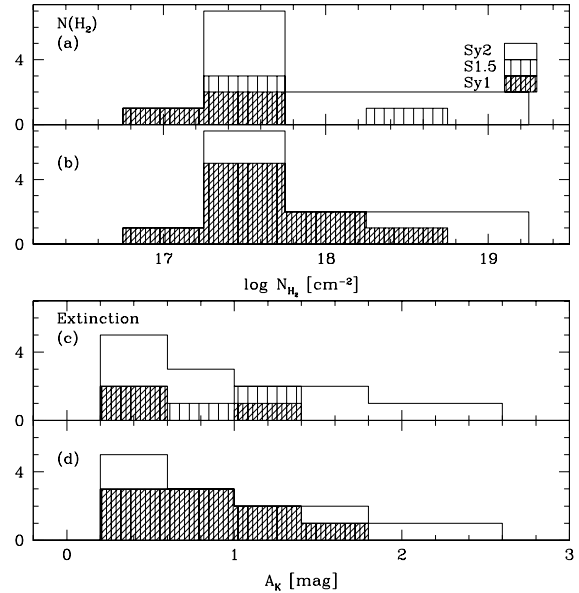}
  \caption{The nuclear surface density of molecular hydrogen ({\em a},
    {\it b}) and the extinction in the nuclear aperture ({\em c}, {\em
      d}). The classification in panels {\em a} and {\em c} is based
    on literature, and in {\it b} and {\it d} on the detection of
    broad Br$\gamma$ (This paper).
\label{histoh2}}
\end{figure}

The only galaxy in the sample with no nuclear {\brg} 2.166 $\mu$m
emission is NGC 1097 (Paper I), which we, however, classify in the
following discussion as Sy1 based on the literature. Broad nuclear
{\brg} emission is detected in 8/14 (57 \%) galaxies in the sample. Of
these galaxies, only two were optically classified as Sy1s and two as
intermediate Seyferts, while the remaining four (NGC 1386, NGC 2110,
NGC 2992 and NGC 7582) were classified as Sy2s. Two of these galaxies
(NGC 1386 and NGC 7582) have no indication of a hidden BLR in
polarized light (Tran 2001). The only other galaxy common with Tran's
sample is NGC 1068, where BLR region is detected in polarized emission
but not in {\brg}. The detection of broad Br$\gamma$ in $\sim$40 \%
(4/9) of the optically classified Sy2s is mainly due to the lower
extinction that the nuclear emission suffers in the NIR (Fig
\ref{histoh2}). The high detection rate of broad Br$\gamma$ is also a
consequence of the high spatial resolution and quality of the data
used.

If the classification is based on the detection of {\brg}, Sy2
galaxies have larger extinction than Sy1s (Fig. \ref{histoh2}).  The
two Sy2s with low $A_{\rmn{K}}$ are ESO 428-G14 and NGC 3081, both of
which are among the least luminous in the sample, while the three high
$A_{\rmn{K}}$ Sy2 galaxies (NGC 1068, NGC 4945 and NGC 5128) represent
the high luminosity end of the sample. Whether the obscuring material
is related to the molecular torus or a foreground dust screen is
unknown; the latter explanation seems more plausible as the optical
depth of the molecular torus is likely to be high enough to completely
obscure the BLR.  The foreground absorbing material may be in the form
of star forming clouds, which are known frequently to coexist with the
AGN (e.g. Storchi-Bergmann et al.  2001). This explanation seems
likely, as 75 \% (6/8) of the galaxies which have broad {\brg} also
have a narrow component. Alternatively narrow {\brg} may arise from
the NLR.

Additionally, in two galaxies (Sy1 NGC 3783 and Sy2 NGC 2992) broad
{\he} 2.058 $\mu$m emission was detected. In both of them,
the width of {\he} is comparable to that of {\brg}, and the broad
dereddened \he/{\brg} flux ratio $\sim$0.35. One possible explanation
for the non-detection of broad {\he} in most of the sample galaxies
is the intrinsic weakness of {\he} with respect to Br$\gamma$,
or the atmospherically poor region that {\he} is located in. The role
of extinction seems unlikely as $A_{\rmn{HeI}}$ $\simeq$
1.1$A_{\rmn{Br\gamma}}$.

\begin{figure}
  \includegraphics[width=8.5cm]{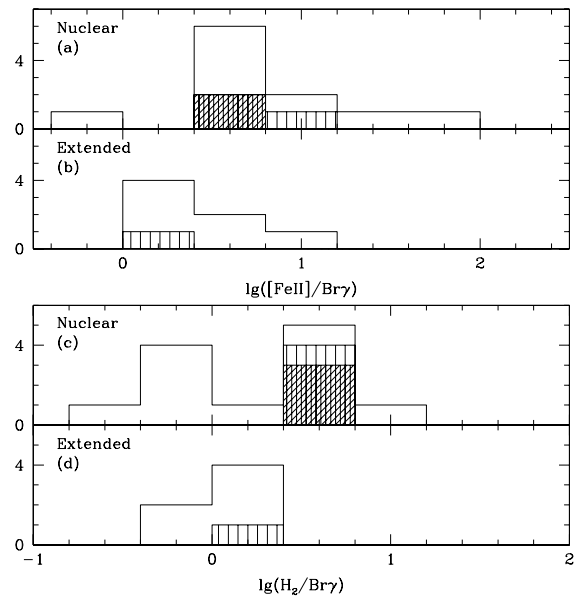}
  \caption{The logarithmic {\iron/\brg$_{narrow}$} ratios
    ({\em a,b}) in the nuclear aperture and for the extended emission,
    and the {\h2/\brg$_{narrow}$} ratios ({\em c,d}). The shading is
    as in Fig.  \ref{histoh2}.
\label{histofe}}
\end{figure}

\subsection{{\iron} emission}

Spatially resolved {\em nuclear} {\iron} 1.644 $\mu$m emission is
detected in 57 \% of the galaxies (8/14; one Sy1, two intermediate
Seyferts, and five Sy2s).  In spatially resolved cases, the FWHM size
of the nuclear emission corrected for the seeing ranges from 40
pc in NGC 3227 to $\sim$300 pc in NGC 7582. 
The total extent of the nuclear {\iron} emission ranges from $\sim$30
pc in NGC 5128 to 1.2 kpc in NGC 2110, with average extent $\sim$300
pc in the plane of the galaxy. In many galaxies {\em separate}
emission regions are detected further out from the nucleus. This
extended emission is very patchy, and in most galaxies closely follows
the star forming complexes as {\iron} is fairly correlated with {\brg}
2.166 $\mu$m and {\he} 2.058 $\mu$m, which both are tracers of star
formation. The origin of this extended {\iron} emission is thus very
likely due to star forming processes and supernovae. The low
{\iron/\brg} ratios 0.5 -- 2 (Fig. \ref{histofe}) of this extended
emission are also generally in agreement with theoretical models of
star forming regions (e.g. Colina 1993).

The origin of the {\em nuclear} {\iron} is however unclear.  Recently,
Rodriguez-Ardila et al. (2002) studied four narrow-line Sy1 galaxies
in the wavelength range 0.8 - 2.4 $\mu$m, and suggested that a
combination of collisional excitation and decays from levels populated
by Ly photons produce the NIR {\iron} spectrum. Other suggestions
include shocks and X-ray excitation.

The weaker $H$-band {\iron} lines are too faint to yield reliable line
ratios, but the \iron/{\brg} ratio is sensitive to the excitation
mechanism.  However, the {\brg} fluxes used in this ratio should
exclude the broad component, because the BLR is unlikely to contribute
much to the {\iron} flux. The {\iron}/Br$\gamma_{narrow}$ ratios are
high in all the galaxies with a broad component, ranging from $\sim$3
in NGC 3783 to $\sim$80 in NGC 2110 (Fig. \ref{histofe}). These ratios
are much higher than star formation models predict, but can be
produced by X-ray or shock excitation. Often, the {\iron} is broader
than the instrumental profile, reaching up to FWHM $\sim$700 {\kms}, and
blueshifted with respect of {\brg}. Both of these are signatures of shock
excitation. In NGC 1386 and NGC 2110 the spatial correlation between
X-ray and {\iron} emission favours X-ray excitation.
In galaxies without broad {\brg} the \iron/{\brg} ratio is also
fairly high, $\sim$3. Only in NGC 3081 this ratio is low, 0.6,
obtainable with shocks, photoionization or star formation. Thus the
{\iron} is predominantly excited by shocks produced by winds/jets in
the ionization cone.

\subsection{\h2 emission}

Nuclear {\h2} emission was detected in all galaxies except in NGC
1365.  The surface densities of the excited {\h2} (Table \ref{prm})
range from $1.5\times10^{17}$ cm$^{-2}$ in NGC 1097 to $1.3\times10^{19}$
cm$^{-2}$ in NGC 1068. The surface densities tend to be larger in Sy2s
than in Sy1s 
(Fig. \ref{histoh2}), but the scatter is large. Also, when
classification into Seyfert types is based on the
detection/non-detection of broad Br$\gamma$ emission, this relation
disappears. This suggests that the optical classification into Seyfert
types is largely related to the extinction effects. However, the
scatter of $N_{\rmn{H_2}}$ is larger in narrow {\brg} galaxies than in
broad {\brg} galaxies.

\begin{figure}
  \includegraphics[width=8.5cm]{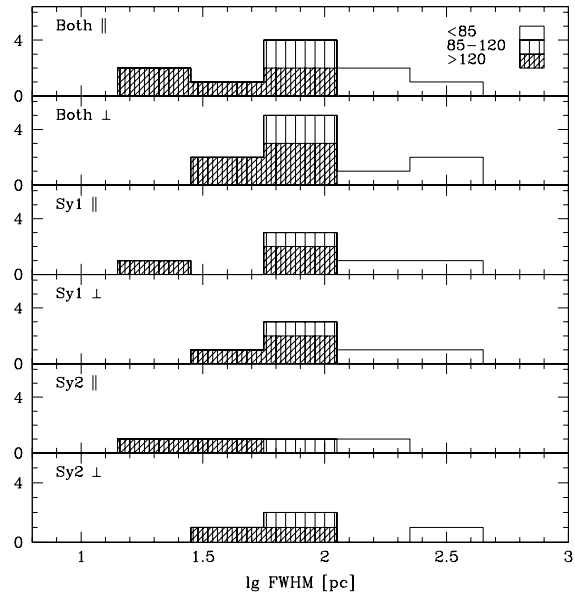}
  \caption{The FWHM size of the \h2 emission both parallel to the cone
  and perpendicular to it for the whole sample ({\em upper two
  panels}) and separatly for both Sy types ({\em lower panels}). The
  classification is based on the detection of broad \brg. Different
  physical scales (pc/arcsec) are indicated with shading.
 \label{histofwhm}}
\end{figure}

The {\em nuclear} 1--0 S(1) 2.122 $\mu$m emission is spatially
resolved in all the galaxies except parallel to the cone in NGC 1386.
The FWHM extent of nuclear H$_2$ emission is larger perpendicular to the
cone than parallel to it in 6/8 (75 \%) galaxies (Fig.
\ref{histofwhm}), while in four galaxies the nuclear H$_2$ emission is
too weak to obtain accurate FWHM sizes and in two galaxies the H$_2$
emission center is significantly off-nuclear. The FWHM sizes corrected
for the seeing range from $<$20 pc in NGC 1386 to $\sim$300 pc in NGC
2110, and are larger then the predicted sizes for the molecular torus
(1-100 pc; Pier \& Krolik 1993). Thus the emission is more likely to
arise from the molecular material (disk?) surrounding the torus than
directly from the torus itself. 

{\em Extended nuclear} emission is detected in all the galaxies except
in NGC 1386 and NGC 3783.  The overall 1--0 S(1) emission is more
extended perpendicular to the cone than parallel to it in five
galaxies, while the opposite is true in seven galaxies.  The extended
nuclear 1--0 S(1) emission can be detected up to a range of physical
scales in the plane of the galaxy, from $\sim$100 pc in NGC 5128 to
$>$1 kpc in NGC 2992 and NGC 4945.  Since this extended nuclear
emission declines smoothly with increasing radius, it seems probable
that the molecular gas forms a disc surrounding the nucleus.  This is
further supported by the 1--0 S(1) velocity curves, which are
generally well-ordered. A starburst origin for this smooth extended
emission seems excluded as the morphologies of both {\brg} and {\iron}
emission are generally different from that of \h2. In seven galaxies,
{\em separate} off-nuclear 1--0 S(1) emission knots are also detected.
In most cases these H$_2$ patches are also detected in {\brg} and
{\iron} and likely originate from starforming regions. In conclusion,
extended {\h2} emission appears to be related with two main components: the
first associated with star forming regions and the second a smoothly
decaying disk, which is not observed in {\iron} or {\brg}.

Only in 36 \% (5/14) of the galaxies, nuclear 2--1 S(1) 2.244 $\mu$m
emission was detected and the 2--1 S(1)/ 1--0 S(1) ratio or their
upper limits are inconsistent with significant fluorescent excitation,
and in agreement with thermal excitation.  Furthermore, off-nuclear
2--1 S(1) emission is detected only in two galaxies, and is also
consistent with thermal excitation of the molecular \h2. This is valid
even in the ring of NGC 1365, where the EW of the {\brg} 2.166 $\mu$m
line is $\sim$70 \AA, indicative of a strong UV radiation field.
Whether the thermal excitation is produced by shocks, X-rays or UV
fluorescence in dense clouds, remains unclear, although shocks are the
most likely explanation.

\subsection{Coronal line emission}

Four coronal lines were detected: {\sivi} 1.964 $\mu$m, {\alix} 2.043
$\mu$m, {\caviii} 2.321 $\mu$m and {\sivii} 2.4833 $\mu$m.  Of these
lines, {\sivi} was detected in eight, {\alix} in two, {\caviii} in
four and {\sivii} in seven galaxies. Overall, at least one coronal
line was detected in 57 \% of the galaxies (8/14; one Sy1, one
intermediate Seyfert and six Sy2s). This substantially increases the
number of Seyferts with IR coronal line detection, and confirms the
results of Prieto \& Viegas (2000), that coronal lines
are common and bright in the spectra of Seyfert galaxies.  The {\alix}
line has been detected so far to our knowledge only in Circinus
(Maiolino et al. 1998), NGC 2992 and NGC 3081 (this paper). In
addition, in a few galaxies either {\caviii} or {\alix} was detected
at $\sim$2.5$\sigma$ level, and awaits confirmation.  Coronal lines
are detected in 2/5 galaxies (40 \%), where {\brg} 2.166 $\mu$m is
narrow, and in 5/8 galaxies (62 \%) where {\brg} is broad.

Interestingly, the coronal lines are extended parallel to the cone in
all three galaxies with spatial information, i.e. NGC 1068, NGC 3081
and ESO 428-G14, implying an anisotropic nuclear radiation field.  In
NGC 1068 the extended emission is only visible on one side, in NGC
3081 both sides are similarly extended and in ESO 428-G14 the emission
is more extended toward southwest than northeast. Furthermore, in NGC
3783 the coronal lines are spatially resolved, but the size is similar
both parallel and perpendicular to the ionization cone. We have an
on-going followup program to obtain diffraction limited images of
nearby Seyferts in the {\sivii} line to study the coronal line region
in detail (Prieto et al. 2003).

\section{Conclusions}

Near-infrared 1.5--2.5 $\mu$m long-slit spectroscopy with a resolution
of $\sim$300 \kms of eight nearby Seyfert galaxies is presented. Together
with similar data of additional six Seyfert galaxies (Paper
I), a homogeous sample of high quality NIR spectra with a spatial
resolution $\leq$1 arcsec is produced both parallel to the ionization
cone and perpendicular to it. This dataset allows us to study the
spatial extension and kinematics of \iron, {\h2}, {\brg} and coronal
lines, and the sizes of the central emitting regions down to $<$0.5
arcsec (10 - 100 pc), and the central molecular content within a 1.4
arcsec diameter aperture. The main results are:

\begin{enumerate}
\item{The strongest NIR emission lines in the nuclear spectra are
    {\brg}, \h2 2.12 $\mu$m, {\iron} 1.64 $\mu$m and coronal lines
    {\sivi} 1.96 $\mu$m and {\sivii} 2.48 $\mu$m.}
\item{Broad {\brg} was detected in $\sim$40 \% of the observed Sy2
    galaxies, stressing the importance of extinction effects and
    classifying AGN in the infrared.  The existence of genuine Sy2s on
    the basis of nondetection of the BLR in polarized emission (Tran
    2001) is inconsistent with our detection of broad {\brg} in two of
    the candidates proposed by the author. We instead believe that
    high spatial resolution NIR spectroscopy will reveal more hidden
    broad line objects in the future.}
\item{Large concentrations of H$_2$ are present in the nucleus
    regardless of the Seyfert type. The nuclear surface densities of
    excited H$_2$ range from 10$^{17}$ to 10$^{19}$ cm$^{-2}$. In
    general, two extended components are detected: a smoothly decaying
    disk and (in some cases) circumnuclear starforming ring. While the
    ring is prominent also in {\iron} and {\brg}, the disk is only
    detected in H$_2$.}
\item{According to the unified models the putative torus collimating
    the radiation from AGN should be evident as more extended nuclear
    emission perpendicular to the cone than parallel to it. This is
    indeed detected in 6/8 (75 \%) galaxies.}
\item{There is a good spatial and kinematical correlation between
    {\brg} and {\iron}. Generally two components are distinguished:
    one related with a circumnuclear ring and a nuclear component.
    Extended {\iron} and {\brg} emission is detected in half of the
    sample. Nuclear {\iron} is in many galaxies blueshifted and the
    width of {\iron} reaches up to 700 {\kms} FWHM, which together
    with high {\iron}/{\brg} ratios suggests shock excitation as a
    dominant excitation mechanism in Seyfert galaxies.}
\item{Coronal lines {\sivi} and {\sivii} are present in $\sim$60 \% of
    the objects and also {\caviii} and {\alix} are detected in some
    cases, confirming the results of Prieto \& Viegas (2000) that
    coronal lines are common and bright features in the spectra of
    Seyfert galaxies regardless of the type of the nucleus. In three
    galaxies the coronal lines are extended parallel to the cone but
    not perpendicular to it.  This clearly supports the presence of a
    strong anisotropic radiation field in these galaxies. However, due
    to the high ionization potential of the coronal lines, the
    extended emission cannot be produced by the nuclear ionizing
    radiation unless it is strongly collimated. Thus shock excitation
    caused by e.g. the jet interacting with the interstellar medium is
    the most plausible origin for the extended coronal emission. The
    width of the coronal lines (up to 1200 {\kms}) also supports this
    scenario.}
\end{enumerate}

\section{Acknowledgements}

Based on observations collected at the European Southern Observatory,
La Silla, Chile.  This research has made use of the NASA/IPAC
Extragalactic Database (NED), which is operated by the Jet Propulsion
Laboratory, California Institute of Technology, under contract with
the National Aeronautics and Space Administration. JR acknowledges a
grant from Finnish Cultural Foundation during this work.
\\

\label{lastpage}
\end{document}